\documentclass{imsart} 

\usepackage{amsmath,amssymb,amsthm}
\usepackage{graphicx}
\usepackage{bbm}

\usepackage{hyperref}
\hypersetup{pdftoolbar=true,
	pdfmenubar=true,
	pdfpagemode=UseOutlines,
	bookmarksnumbered=true,
	linktocpage=false,
	colorlinks=false,
	hidelinks
}

\usepackage{bm}
\usepackage{xcolor}
\usepackage[plain,noend]{algorithm2e}

\usepackage{pdfpages} 

\makeatletter
\let\c@author\relax
\makeatother

\usepackage[backend=bibtex8, bibencoding=latin1, style=authoryear-comp, giveninits=true, uniquename=init, dashed=false, useprefix=true, natbib=true, maxcitenames=2, maxbibnames = 9, eprint = true]{biblatex}
\DeclareNameAlias{sortname}{family-given}
\renewbibmacro{in:}{%
	\ifboolexpr{%
		test {\ifentrytype{article}}%
		or
		test {\ifentrytype{inproceedings}}%
		or
		test {\ifentrytype{inbook}}%
	}{}{\printtext{\bibstring{in}\intitlepunct}}%
}

\usepackage{soul}

\bibliography{literature_discrete, literature_general, literature_continuous}

\theoremstyle{plain} 
\newtheorem{theorem}{Theorem}
\numberwithin{theorem}{section}
\newtheorem{corollary}[theorem]{Corollary}
\newtheorem{lemma}[theorem]{Lemma}
\newtheorem{proposition}[theorem]{Proposition}

\theoremstyle{remark}
\newtheorem{example}[theorem]{Example}
\newtheorem{definition}[theorem]{Definition}
\newtheorem{remark}[theorem]{Remark}

\def\N{\mathbb{N}}
\newcommand{\Nz}{\N_{\geq 2}}	
\def\Z{\mathbb{Z}}
\def\R{\mathbb{R}}
\def\supp{\operatorname{supp}}
\def\int{\operatorname{int}}

\newcommand{\oneto}[1]{\{1, \ldots, {#1}\}}	

\def\Rdists{\mathcal{P}}
\def\Qdists{\mathcal{Q}}

\def\Ddists{\mathcal{D}}
\def\PDdists{\mathcal{D}_0}

\def\E{\mathbb{E}}
\def\Prob{\mathbb{P}}

\newcommand{\Intdists}[1]{\mathcal{L}^{#1}}	

\def\dispo{\leq_{\text{disp}}}
\def\dilo{\leq_{\text{dil}}}
\def\gendo{\leq_{\text{D}}}

\def\Unif{\mathcal{U}}
\newcommand{\Bin}{\mathrm{Bin}}			
\newcommand{\Geom}{\mathrm{Geom}}

\newcommand{\Lm}[1]{\lambda^{#1}}	

\newcommand{\sto}{\leq_{\text{st}}}		
\newcommand{\dispoeq}{=_{\text{disp}}}		
\newcommand{\discoa}{\leq_{\text{disp}}^{\land\text{-disc}}}	
\newcommand{\discoaeq}{=_{\text{disp}}^{\land\text{-disc}}}	
\newcommand{\discoo}{\leq_{\text{disp}}^{\lor\text{-disc}}}	
\newcommand{\discooeq}{=_{\text{disp}}^{\lor\text{-disc}}}	

\newcommand{\cdfis}{\; \hat{=} \;}		

\newcommand{\compj}[2]{\stackrel{{\scriptscriptstyle {#1}, {#2}}}{\leftrightharpoons}}	
\newcommand{\stdCompj}{\leftrightharpoons}	
\newcommand{\compja}[2]{\stackrel{{\scriptscriptstyle {#1}, {#2}}}{\leftrightharpoons_\land}}	
\newcommand{\stdCompja}{\leftrightharpoons_\land}	
\newcommand{\compjo}[2]{\stackrel{{\scriptscriptstyle {#1}, {#2}}}{\leftrightharpoons_\lor}}	
\newcommand{\stdCompjo}{\leftrightharpoons_\lor}	

\newcommand{\indsetset}{\mathbb{I}}		
\newcommand{\idnseqset}{\mathbb{S}}		

\newcommand{\smi}[1]{\underline{#1}}	
\newcommand{\sma}[1]{\overline{#1}}		

\newcommand{\nn}[3]{\operatorname{NN}_{#2}^{#3}(#1)}	

\newcommand{\sd}{\mathrm{SD}}		
\newcommand{\gmd}{\mathrm{GMD}}		
\newcommand{\mad}{\mathrm{MAD}}		
\newcommand{\mdmad}{\mathrm{MDMAD}}	
\newcommand{\iqnr}[2]{\mathrm{IQR}({#1}, {#2})}	
\newcommand{\iqr}{\mathrm{IQR}}		
\newcommand{\ienr}[2]{\mathrm{IER}({#1}, {#2})}	

\allowdisplaybreaks

\begin{document}

\begin{frontmatter} 
\title{Defining Dispersion: A Fundamental Order for Univariate Discrete Distributions}
\runtitle{Defining Dispersion: A Fundamental Order for Univariate Discrete Distributions} 

\begin{aug}
\author{\fnms{Andreas} \snm{Eberl}\ead[label=e1]{andreas.eberl@kit.edu}}
\address{Institute of Statistics \\
    Karlsruhe Institute of Technology (KIT) \\
    \printead{e1}}

\and
\author{\fnms{Bernhard} \snm{Klar}\ead[label=e2]{bernhard.klar@kit.edu}}
\address{Institute of Stochastics \\
    Karlsruhe Institute of Technology (KIT) \\
    \printead{e2}}

\end{aug}



\begin{abstract}
    The measurement of dispersion is one of the most fundamental and ubiquitous statistical concepts, in both applied and theoretical contexts. For dispersion measures, such as the standard deviation, to effectively capture the variability of a given distribution, they must, by definition, preserve some stochastic order of dispersion. The so-called dispersive order is the most basic order that serves as a foundation underneath the concept of dispersion measures. However, this order is incompatible with almost all discrete distributions, including lattice and most empirical distributions. As a result, popular measures may fail to accurately capture the dispersion of such distributions.     

    In this paper, discrete adaptations of the dispersive order are defined and analyzed. They are shown to be a compromise between being equivalent to the original dispersive order on their joint area of applicability and other crucial properties. 
    Moreover, they share many characteristic properties with the dispersive order, validating their role as a foundation for measuring discrete dispersion in a manner closely aligned with the continuous setting.      
    Their behaviour on well-known families of lattice distribution is generally as expected when parameter differences are sufficiently large. Most popular dispersion measures preserve both discrete dispersive orders, rigorously ensuring that they are also meaningful in discrete settings. However, the interquantile range fails to preserve either discrete order, indicating that it is unsuitable for measuring the dispersion of discrete distributions.
\end{abstract}

\end{frontmatter}


\tableofcontents

\section{Introduction}
\label{sec:intro}

The measurement of dispersion is one of the fundamental concepts in statistics, playing a central role in both theoretical and applied contexts. It is widely used in mathematical statistics, applied sciences, and empirical research. While measures of dispersion have been in use since at least the early 19th century \citep{variance_hist}, their first rigorous definition was introduced by \citet{bl3, bl4}.  

Since the systematic study of statistical measures of characteristics such as location, dispersion, and skewness began, it has been widely agreed that these measures should align with an appropriate stochastic order. The seminal work of \citet{bl2} provides several arguments supporting the requirement that a measure of central location, $\nu$, be consistent with the usual stochastic order. This means that if $F \leq_{st} G$ for cumulative distribution functions (cdf's) $F$ and $G$, then $\nu(F) \leq \nu(G)$. Additionally, $\nu$ must satisfy the affine linear transformation property $\nu(aX+b) = a\nu(X) + b$ for all $a,b\in \mathbb{R}$ for a random variable $X$.  

Similarly, \citet{bl3, bl4} proposed two essential properties for a dispersion measure, \(\tau\). The first property stipulates its behavior under affine linear transformations: \(\tau(aX+b) = |a|\tau(X)\) for all \(a, b \in \mathbb{R}\). However, as discussed in Section \ref{sec:basics}, this property alone does not yield meaningful measures of dispersion.  

The second property is more critical: \(\tau\) must preserve a given dispersion order, \(\gendo\), meaning that if \(F \gendo G\), then \(\tau(F) \leq \tau(G)\). 
The ordering $F \gendo G$ must be defined in such a way that it reflects the greater concentration of $F$ compared with $G$ throughout the support of $F$.
This second property is fundamental to the meaningful measurement of dispersion, as it links the measure to the concept of one distribution being more dispersed than another through \(\gendo\). Consequently, the choice of \(\gendo\) serves as the conceptual foundation for understanding and defining dispersion.
 
The dispersive order $\dispo$ is the strongest order of dispersion in the literature and therefore imposes the most basic requirement on dispersion measures. The order requires one distribution to be more spread out than another throughout their entire quantile functions in a pointwise sense. Other popular, but weaker, dispersion orders like the dilation order $\dilo$ not only compare two distributions using aggregated quantities but also inherently favour some structurally similar dispersion measures. In the case of the dilation order, the distributions are centred around the mean and therefore dispersion measures like the standard deviation are given an advantage. Thus, the dispersive order is usually chosen as the fundamental order for dispersion measures in the literature (see, e.g., \citealp{bl4} and \citealp{oja}).

A critical limitation of the dispersive order is revealed in a seemingly innocuous result by \citet[Th. 1.7.3]{mueller-stoyan}, which has gone largely unnoticed in the literature. This result states that a necessary condition for $F \dispo G$ is that the range of $F$ must be a subset of the range of $G$. While this condition poses no issue for continuous distributions, it effectively excludes almost all classes of discrete distributions from being ordered using $\dispo$. Among the excluded classes are nearly all lattice distributions and all empirical distributions that either exhibit ties or differ in sample size. This fact has been completely ignored in the literature. For instance, in the authoritative compendium on stochastic orders by \citet{shaked-sh}, the dispersive order is defined for general distributions and analyzed extensively without any mention of this critical restriction. Consequently, the results discussed are inapplicable to discrete distributions.  
This stands in stark contrast to the usual stochastic location order, which can be directly applied to discrete distributions without modification, and undermines the foundation of dispersion measures for discrete distributions -- a foundation that has been assumed valid for decades.  

An obvious example is given by two discrete uniform distributions, one on the set $\{1, 2\}$ and the other on the set $\{1, \ldots, 5\}$. Although it is self-evident that the second distribution is more dispersed, the dispersive order does not come to that conclusion. Despite examples like this, using measures to quantify the dispersion of lattice distributions and empirical distributions is often among the first topics in an introductory statistics course. It is commonplace in all kinds of applied sciences. The field of mathematical statistics must ensure that a concept this widely used is well-founded and meaningful.

Therefore, this article's goal is not to introduce yet another dispersion order among the many that already exist. Instead, it seeks to establish a fundamental dispersion order specifically for univariate discrete distributions. This new order transfers many of the properties of the dispersive order, as applied to continuous distributions, to discrete distributions. In doing so, it seeks to reconstruct the foundational framework for measuring dispersion in discrete distributions.

Such a foundational order is not only of theoretical interest but also has direct implications for other statistical theories and practical tools. For example, in a forthcoming paper, \citet{eks-weakDisp} propose a stochastic order for discrete distributions based on the L\'evy concentration function, which is closely related to the theory of majorization. By construction, this can be interpreted as a concentration order. Among other contributions, they introduce a robust measure derived from this order. Since they further demonstrate that this new order is consistent with the discrete dispersive order developed in the present work, it is fully justifiable to regard their order as a dispersion order and the associated measure as a robust dispersion measure - something that appears to be rather rare for discrete models.

Following this introduction, the dispersive order is formally introduced and motivated. We then point out its shortcomings for discrete distributions in detail with the help of an illustrative example. Section \ref{sec:basics} also contains two real data examples in which the dispersive order fails.

In the following Section \ref{sec:discDispDef}, two possible discrete versions of the dispersive order are defined. Either of the two orders satisfies all but one crucial properties that one would hope for, see Section \ref{sec:discDispDef}. Therefore, it is unclear which order is superior and both are considered throughout the remainder of the article. A heuristic explanation of why we conjecture that there exists no order that satisfies all of the aforementioned crucial properties is given at the end of Section \ref{sec:heuristics}. Beforehand, several properties of the original dispersive order are replicated for the two discrete versions and some supplementary results and examples give insights into the derivation and the interpretation of the discrete orders.

The then following Section \ref{sec:discDispMeas} is of particular interest as it draws the connection from the world of stochastic orders back to the more application-relevant concept of dispersion measures. Most popular dispersion measures like the standard deviation, the mean absolute deviation from the mean and Gini's mean difference preserve both discrete dispersive orders and therefore measure the dispersion of discrete distributions in a meaningful way. All three results are corollaries of the fact that the dilation order is a weakening of the discrete dispersive orders for discrete distribution; the same is true for the original dispersive order in a continuous setting. Unsurprisingly, the dilation order is not helpful for quantile-based dispersion measures like the mean absolute deviation from the median or interquantile distances. While the mean absolute deviation from the median can still be shown to preserve the discrete dispersive orders, interquantile distances generally do not have this property.

This reveals a substantial problem since the interquartile range is a popular measure in descriptive statistics. Using it in the context of empirical or lattice distributions, especially with small supports, can lead to fundamental misunderstandings concerning the given data set. Combined with the fact that the counterexamples used to prove this result are rather simple, this provides a convincing and rigorous argument against using this measure in a discrete setting. This incompatibility has previously been noted in the literature, most recently by \citet[p.\ 1852]{IER}, who deemed the interquantile range not to be a ``true measure of variability''. However, the given reasoning is that it does not preserve the dilation order, which is not sufficient to infer this kind of statement. For continuous distributions, the interquartile range is indeed a true measure of variability. Furthermore, the fact that it does not preserve the discrete dispersive orders is not dependent upon the used definition of quantiles, which is not unique for discrete distributions.

The derived discrete dispersive orders are applied to several popular discrete distributions in Section \ref{sec:specDists} to find out whether they are systematically ordered in the seemingly evident direction. This is the case, if the parameter difference within the distribution family is large enough. For smaller parameter differences, the situation is less clear, which can be explained by the sparse nature of discrete distributions.

All proofs as well as applications to further popular discrete distributions are deferred to the supplement.

\section{The classical foundation of dispersion measurement and its shortcomings for discrete distributions}
\label{sec:basics}

Let $\Rdists$ denote the set of all real-valued probability distributions and let $\Qdists \subset \Rdists$ be a suitable subset. We allow $\Rdists$ and subsets thereof to be interpreted as sets of cumulative distribution functions (cdf's). For each cdf $F \in \Rdists$, we furthermore define the interval
\begin{equation*}
    D_F = \R \setminus F^{-1}(\{0, 1\}) = \{t \in \R: F(t) \in (0, 1)\}.
\end{equation*}
For example, if $F$ represents a normal distribution, then $D_F = \mathbb{R}$; if $F$ is a beta distribution, then $D_F = (0, 1)$; if $F$ is a binomial distribution, then $D_F = [0, n)$; and if $F$ is a Poisson distribution, then $D_F = [0, \infty)$. Sets of this kind were also studied in the pioneering works of \citet[p.\ 6]{zwet}, who denoted them as $I$, and \citet[p.\ 155]{oja}, who denoted them as $S_F$. When cdf's $F$ and $G$ are considered, the corresponding random variables are $X \sim F$ and $Y \sim G$, respectively.

Throughout the paper, we call a distribution discrete if its support is at most countable, and we call a distribution continuous if it is absolutely continuous with respect to the Lebesgue measure.

Given an agreed-upon order of dispersion $\gendo$, the classical definition of a dispersion measure due to \citet{bl3} is given as follows.

\begin{definition}
\label{def:dispMeas}
    A mapping $\tau: \Qdists \to [0, \infty)$ is said to be a {\em measure of dispersion}, if the following two properties are satisfied:
    \begin{enumerate}
        \item[(D1)] $\tau(aX+b) = |a| \tau(X)$ for all $a, b \in \R$ and $F \in \Qdists$,
        \item[(D2)] $\tau(F) \leq \tau(G)$ for all $F, G \in \Qdists$ with $F \gendo G$.
    \end{enumerate}
\end{definition}

The two most popular dispersion orders in the literature are defined as follows \citep[see, e.g.,][]{mueller-stoyan}.

\begin{definition}
\label{def:dispoDilo}
    Let $F, G \in \Rdists$.
    \begin{enumerate}
        \item[a)] $F$ is said to precede $G$ in the {\em dispersive order}, in short $F \dispo G$, if
        \begin{equation*}
            F^{-1}(p_1) - F^{-1}(p_0) \leq G^{-1}(p_1) - G^{-1}(p_0) \quad \forall 0 < p_0 \leq p_1 < 1.
        \end{equation*}

        \item[b)] Let the first moments of $F$ and $G$ exist. Then, $F$ is said to precede $G$ in the {\em dilation order}, in short $F \dilo G$, if
        \begin{equation*}
            \E\left[\varphi(X - \E[X])\right] \leq \E\left[\varphi(Y - \E[Y])\right]
        \end{equation*}
        for all convex functions $\varphi$ for which the expectations exist.
    \end{enumerate}
\end{definition}

\begin{remark} \label{rem:D2} 
\begin{enumerate}
\item[(i)]
Condition (D1) in Definition \ref{def:dispMeas} imposes restrictions on the values of the dispersion measure for linearly related distributions and particularly ensures that $\tau$ measures dispersion in a symmetric way. 
Sometimes, (D1) alone is used to define a measure of scale (e.g., in \cite{gerstenberger}). However, there are several compelling reasons why this requirement alone is insufficient to define a measure of dispersion:
\begin{itemize}
\item
There exist many measures that satisfy (D1) but cannot be reasonably considered measures of dispersion. Examples include $|\text{Median} - \text{Mean}|$ and $Q_{3/4}-2Q_{1/2}+Q_{1/4}$, where $Q_p$ denotes the $p$-quantile. These are, in fact, measures of skewness, or more specifically, the numerators of skewness measures (with the denominators scaled by a dispersion measure). Thus, using them as measures of dispersion is nonsensical.  
\item 
A reasonable requirement for a dispersion measure for absolutely continuous distributions is that $X+h(X)$, where $h$ is an increasing function, should have greater dispersion than $X$.
While (D1) ensures this only for $h(X) = X$, condition (D2), in conjunction with the dispersive order, ensures this for any monotonically increasing function $h$.
\item 
Consider the dispersion measure $\tau = 1/f(Q_{1/2})$, where $f$ is the density function of a random variable $X$. This satisfies (D2) and is a reasonable measure of dispersion for absolutely continuous distributions. At first glance, the measure $\tau = 1/f(E[X])$ appears equally valid. However, one can find a positive random variable $X$ such that $\tau(X + X^2) < \tau(X)$, violating the intuitive expectation of increasing dispersion. This discrepancy occurs because $\tau$ does not satisfy (D2). Thus, condition (D2) is crucial for distinguishing between measures that may seem similar in construction but differ fundamentally in their behavior.  
\end{itemize}
\item [(ii)]
As a consequence, (D2) is the crucial condition in Definition \ref{def:dispMeas}. The chosen dispersion order $\gendo$ serves as the foundation for dispersion measures, ensuring their meaningfulness.  
\item[(iii)] 
As mentioned in the introduction, similar considerations apply to measures of location and skewness. An example of counterintuitive behavior arising from measures that fail to preserve a corresponding order is discussed in the context of skewness by \citet{ek-patil}. This example underscores the importance of an order-based definition.  
\end{enumerate}
\end{remark}


The dispersive order can be equivalently characterized using Q-Q-plots with support of $F$ and $G$ on the x- and y-axis, respectively. 
$F \dispo G$ holds, if and only if any straight line connecting two points in the corresponding Q-Q-plot has a slope of at least one.

The dilation order is closely related to the more well-known convex order, which is defined in the same way without the standardization with respect to the mean. The following proposition contains two helpful properties of the two orders \citep[see, e.g.,][]{mueller-stoyan}.

\begin{proposition}
\label{thm:diloProps}
    Let $F, G \in \Rdists$.
    \begin{enumerate}
        \item[a)] $F \dilo G$ is equivalent to $\pi_{X - \E[X]}(t) \leq \pi_{Y - \E[Y]}(t)$ for all $t \in \R$, where $\pi_Z: \R \to [0, \infty), t \mapsto \E[(Z-t)_+]$ denotes the {\em stop-loss transform} of a random variable $Z$.

        \item[b)] $F \dispo G$ implies $F \dilo G$.
    \end{enumerate}
\end{proposition}

For the definition of dispersion measures as given in Definition \ref{def:dispMeas}, the dispersive order $\dispo$ is usually chosen as the foundational order (see, e.g., \citealp{bl4} and \citealp{oja}). In addition to the arguments given in Remark \ref{rem:D2}, one major reason for this choice is that $\dispo$ is the strongest order of dispersion that is commonly considered in the literature; in particular, it is stronger that the dilation order $\dilo$. While $F \dispo G$ means that $G$ is more dispersed than $F$ in a pointwise fashion, $F \dilo G$ represents a comparison of averages that can, e.g., be expressed via the stop-loss transform. Thus, $\dispo$ imposes a minimal requirement on the notion of a dispersion measure. If the dispersive order deems one distribution to be more dispersed than another, this statement is strong enough that every reasonable dispersion measure must share this preference. A second major reason is that the dispersive order does not inherently center the considered distributions around some measure of central location and therefore does not inherently prefer certain dispersion measures to others. For two distributions to be ordered with respect to $\dispo$, one must spread out more than the other in a pointwise sense. On the other hand, the dilation order compares the dispersion of two distributions by first centering them around the mean, thus inherently favouring corresponding dispersion measures like the standard deviation. So, when referring to measures of dispersion or dispersion measures throughout this paper, we mean mappings as defined by Definition \ref{def:dispMeas} with $\dispo$ in the role of $\gendo$.

The following result concerning the dispersive order is given by \citet[p.\ 41]{mueller-stoyan}. Note that $F(D_F)$ is the range of the cdf $F$ excluding $0$ and $1$.

\begin{proposition}
\label{thm:dispoRanges}
	Let $F, G \in \Rdists$. Then, $F \dispo G$ implies $F(D_F) \subseteq G(D_G)$.
\end{proposition}

Conversely, this means that, if neither range of two cdf's is a subset of the range of the other cdf, the two distributions are not ordered with respect to $\dispo$. This does not present an obstacle for continuous distributions since the range of their cdf's always equals the entire unit interval. Discrete cdf's, however, take at most countably many values. This means that, if two discrete ranges were picked at random (via independent uniformly distributed random variables), the probability for the ranges to even coincide in one point would be zero. This problem persists when we consider specific families of distributions like the binomial, Poisson or geometric distributions. 
An exception is given by certain classes of empirical distributions. In particular, every pair of empirical distributions with the same sample size and no ties satisfies $F(D_F)=G(D_G)$, and thus is not a priori excluded from being ordered with respect to $\dispo$.\par
However, it is easy to find examples where one distribution is clearly more dispersed than the other, yet neither \( F(D_F) \subseteq G(D_G) \) nor \( G(D_G) \subseteq F(D_F) \) holds, indicating that the two distributions are not comparable with respect to \( \dispo \). A particularly simple example is provided below.

\begin{example}
\label{exm:2Unif1}
    Let $X \sim \Unif(\{1, 2\})$, $Y \sim \Unif(\{1, \ldots, 5\})$ and $\tilde{Y} \sim \Unif(\{1, \ldots, 4\})$. We have $F(D_F) = \{\tfrac{1}{2}\}$ and $G(D_G) = \{\tfrac{1}{5}, \tfrac{2}{5}, \tfrac{3}{5}, \tfrac{4}{5}\}$, thus, $F \not\dispo G$ (and $G \not\dispo F$). On the other hand, it is easy to show $X \dispo \tilde{Y}$. In particular, this can be seen by observing the corresponding Q-Q-plot in Figure \ref{fig:CEXdispoDisc}.
	
	\begin{figure}
		\centering{
		  \includegraphics[width=\textwidth]{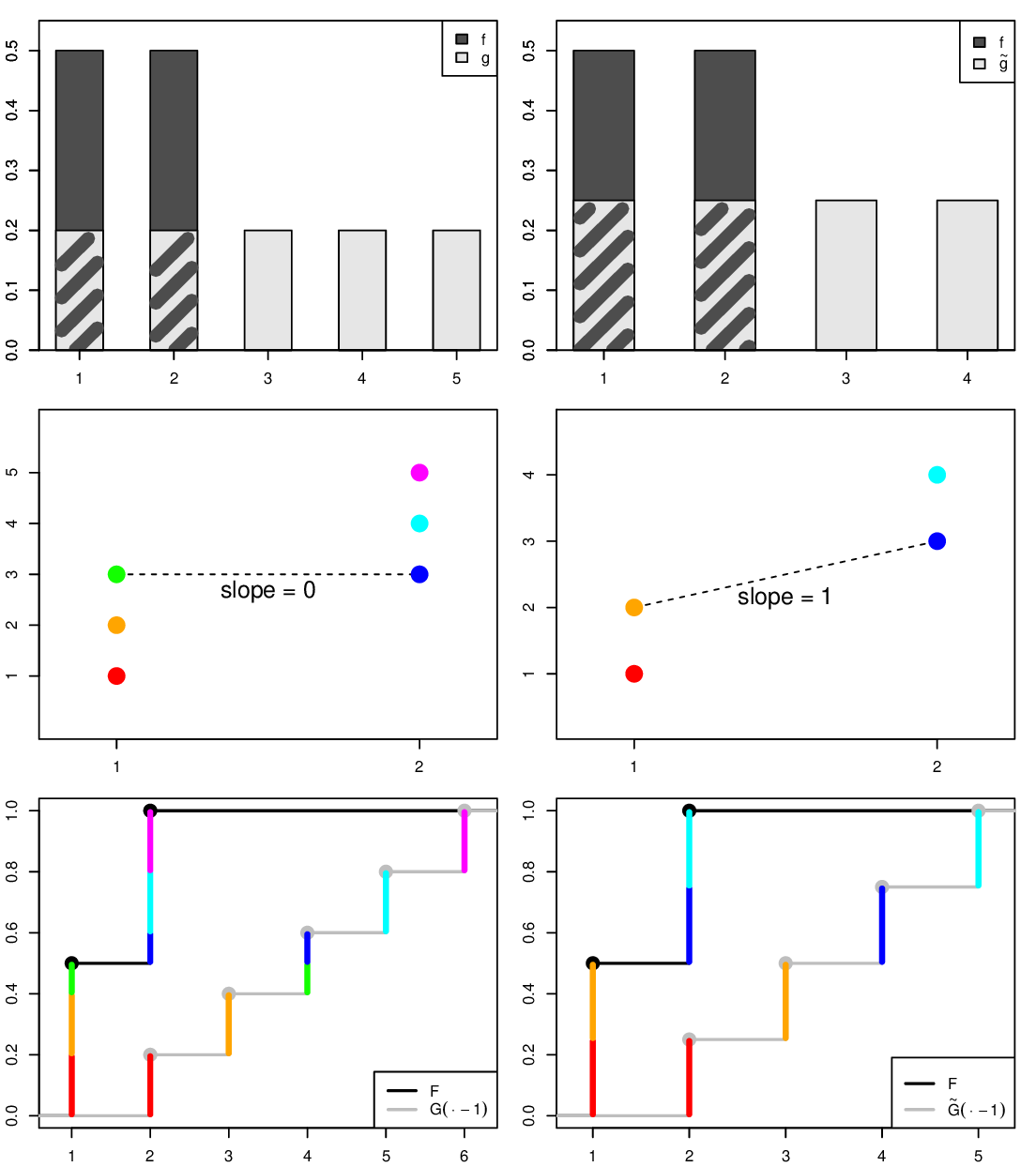}
		  \caption{\label{fig:CEXdispoDisc}Different illustrations of the two pairs of distributions considered in Example \ref{exm:2Unif1}: $X \sim \Unif(\{1, 2\})$ and $Y \sim \Unif(\{1, \ldots, 5\})$ in the left panels; $X$ and $\tilde{Y} \sim \Unif(\{1, 2, 3, 4\})$ in the right panels. Upper: barplots of pmf's; Middle: Q-Q-plots; Lower: colour-coded cdf's.}
		}
	\end{figure}

    The pmf's of $X$ and $Y$ are given in the upper left panel and the pmf's of $X$ and $\tilde{Y}$ are given in the upper right panel. It is obvious that both $X \gendo Y$ and $X \gendo \tilde{Y}$ should hold for any sensible dispersion order $\gendo$. The difference in dispersion between $X$ and $Y$ is even larger than that between $X$ and $\tilde{Y}$, yet $X \not\dispo Y$ and $X \dispo \tilde{Y}$. The reason for this becomes more obvious by observing the corresponding Q-Q-plots in the two middle panels of Figure \ref{fig:CEXdispoDisc}. The critical point in both Q-Q-plots is the jump of $F^{-1}$ from the value $1$ to the value $2$. On the left side, the slope between the two corresponding points is zero, which contradicts $X \dispo Y$; on the right side, the slope between these two points is one, and therefore $X \dispo \tilde{Y}$ is true.\par
    The difference in the two Q-Q-plots can be explained by the comparison of the cdf's, which are depicted in the lower panels of Figure \ref{fig:CEXdispoDisc}. It is colour-coded into the cdf's how the probability mass is shared between the different points in the supports of both distributions. Since this is essentially what is represented by the Q-Q-plots, the colours of the points in the Q-Q-plots correspond to those used for the cdf's. The reason for the slope of zero in the Q-Q-plot on the left side is that a smaller jump of $G$ lies in between two larger jumps of $F$. This is not the case on the right side precisely because the condition $F(D_F) \subseteq G(D_G)$ is fulfilled.
\end{example}

Similar situations can also be observed in real-world data examples like the following.

\begin{example} \label{exm:para1}
We consider two real-world data examples from parasitology. The datasets originate from \citet{Munderle:2005} and are partially described in \citet{Munderle:2006}. Datasets from \citet{Munderle:2005} were also analyzed in \citet{Klar:2010} in the context of the usual stochastic location order for discrete distributions.  

\begin{itemize}
\item [(a)]
%
The first example compares counts of swimbladder-nematode larvae in two populations of the European eel ({\itshape Anguilla anguilla}). The first sample, with a size of $n_1 = 28$, is from the River Sauer near Rosport in Luxembourg. The second sample, with $n_2 = 50$, is from the IJsselmeer near Makkum in the Netherlands. All recorded nematodes belong to the species {\itshape Anguillicoloides crassus}. 

Figure \ref{fig:ex1} displays bar plots of the relative frequencies $p$ and $q$ for the two datasets. Table \ref{tab:ex1} presents the absolute frequencies $h_i (i=1,2)$ alongside the relative frequencies for both datasets.

\begin{table}
\setlength{\tabcolsep}{2pt}
\begin{tabular}{rrrrrrrrrrrrrrrr} \hline
$k$ & 0 & 1 & 2 & 3 & 4 & 5 & 6 & 7 & 9 & 10 & 11 & 12 & 19 & 20 & 48 \\ \hline
$h_1(k)$ & 15 & 5 & 4 & 2 & 2 & 0 & 0 & 0 & 0 & 0 & 0 & 0 & 0 & 0 & 0 \\ 
$p_k$ & 0.536 & 0.179 & 0.143 & 0.071 & 0.071 & 0 & 0 & 0 & 0 & 0 & 0 & 0 & 0 & 0 & 0 \\ \hline
$h_2(k)$ & 17 & 7 & 7 & 3 & 3 & 1 & 4 & 1 & 1 & 1 & 1 & 1 & 1 & 1 & 1 \\ 
$q_k$ & 0.34 & 0.14 & 0.14 & 0.06 & 0.06 & 0.02 & 0.08 & 0.02 & 0.02 & 0.02 & 0.02 & 0.02 & 0.02 & 0.02 & 0.02 \\
\end{tabular}
\caption{\label{tab:ex1} Absolute frequencies $h_1$ and $h_2$ and relative frequencies $p$ and $q$ for the datasets in Example \ref{exm:para1}(a)}
\end{table}

\begin{figure}
\includegraphics[width=\textwidth]{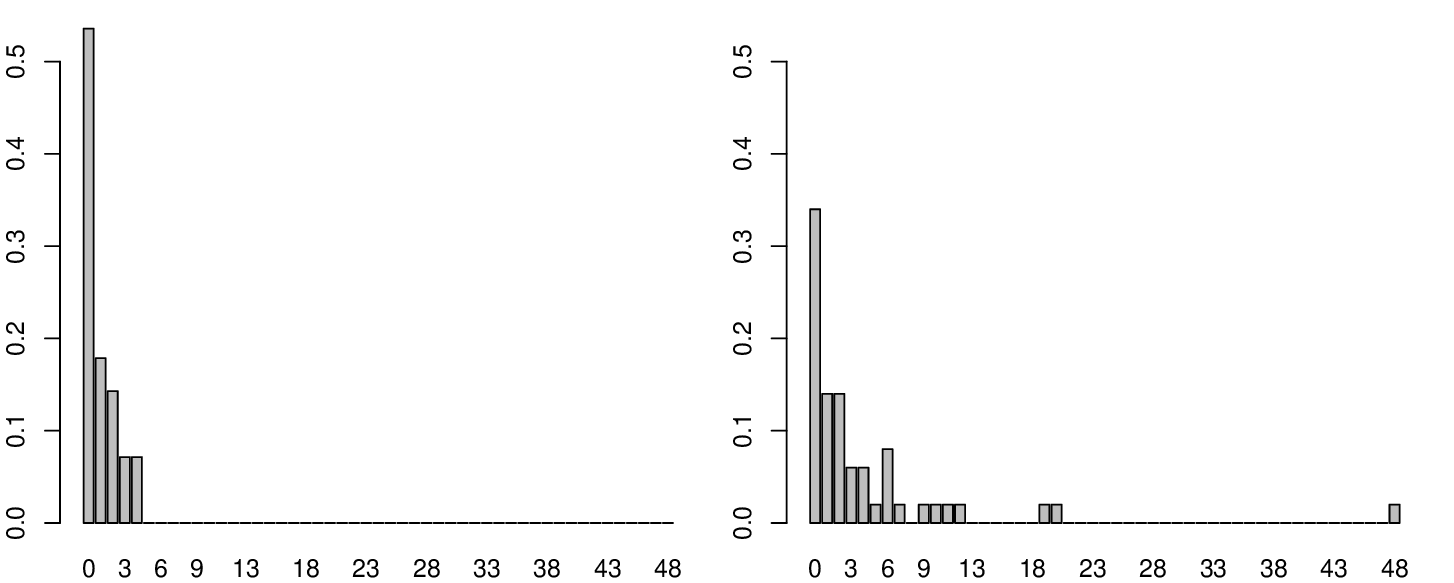}
\caption{\label{fig:ex1} Bar plots of the relative frequencies for the datasets in Example \ref{exm:para1}(a)}
\end{figure}

Note that $p_i > q_i$ for $i = 0, \ldots, 4$, i.e., for each point in the support of $p$. Consequently, $q$ can be derived from $p$ by successively transferring probability mass as follows: 0.011 from point 4 to 5, 0.011 from point 3 to 5 and 6, 0.003 from point 2 to 6, 0.039 from point 1 to 6, and finally 0.196 from point 0 to the points $6, 7, \ldots, 48$. In each step, the spread of the probability mass increases, making it evident that $q$ should exhibit greater dispersion than $p$.
However, the two datasets cannot be ordered with respect to the dispersive order $\dispo$.

\item [(b)]
As the second example, we examine counts of swimbladder-nematode larvae in European eels from two different locations: Sample 1 ($n_1=43$) from the River Rhine near Arnhem in the Netherlands, and sample 2 ($n_2=100$) from the River Rhine near Sulzbach. As before, all recorded nematodes belong to the species Anguillicoloides crassus.

Fig. \ref{fig:ex2} illustrates the relative frequencies $p$ and $q$ of the two datasets. Table \ref{tab:ex2} provides the absolute frequencies $h_i (i=1,2)$ alongside the relative frequencies of both datasets.

\begin{table}[ht]
\setlength{\tabcolsep}{2pt}
\centering
\begin{tabular}{rrrrrrrrrrrrr} \hline
 & 0 & 1 & 2 & 3 & 4 & 5 & 6 & 7 & 8 & 10 & 11 & 12 \\ \hline
$h_1$ & 32 & 8 & 1 & 2 & 0 & 0 & 0 & 0 & 0 & 0 & 0 & 0 \\ 
$p$ & 0.744 & 0.186 & 0.023 & 0.047 & 0 & 0 & 0 & 0 & 0 & 0 & 0 & 0 \\ \hline
$h_2$ & 61 & 16 & 10 & 1 & 2 & 1 & 1 & 1 & 2 & 2 & 2 & 1 \\ 
$q$ & 0.61 & 0.16 & 0.10 & 0.01 & 0.02 & 0.01 & 0.01 & 0.01 & 0.02 & 0.02 & 0.02 & 0.01 \\ \hline
\end{tabular}
\caption{\label{tab:ex2} Absolute frequencies $f_1$ and $f_2$ and relative frequencies $p$ and $q$ for the datasets in Example \ref{exm:para1}(b)}
\end{table}

\begin{figure}
\includegraphics[width=\textwidth]{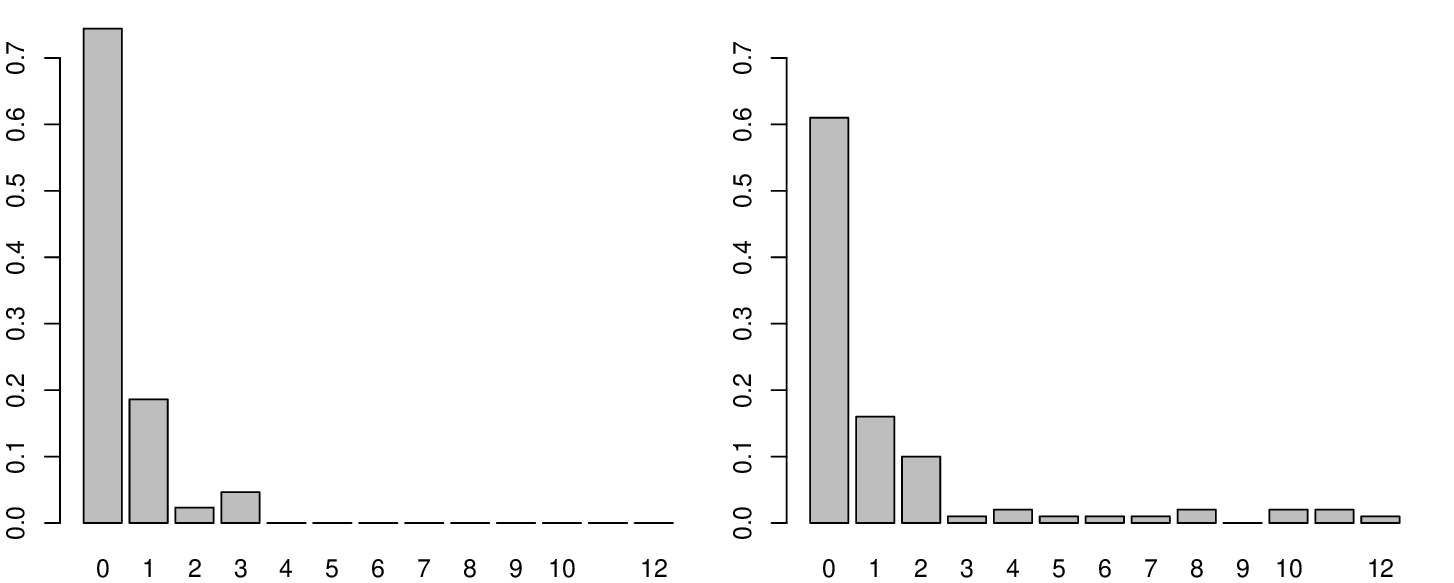}
\caption{\label{fig:ex2} Bar plots of the relative frequencies for the datasets in Example \ref{exm:para1}(b)}
\end{figure}

Looking at Figure \ref{fig:ex2}, most people would likely agree that $q$ is more dispersed than $p$, but the situation is not as clear-cut as in the first example. 
We observe $p_i > q_i$ for $i=0,1,3$, but $p_2<q_2$. Once again, the two datasets cannot be ordered with respect to the dispersive order $\dispo$.
\end{itemize}
\end{example}

The fact that constellations as in Examples \ref{exm:2Unif1} and \ref{exm:para1}a) exist, implies that $\dispo$ does not order discrete distributions sufficiently well with respect to dispersion. Recalling Definition \ref{def:dispMeas}, this leaves the concept of dispersion of discrete distributions without a foundation. In particular, the idea that popular dispersion measures like the standard deviation, the quantile distance and many others actually measure dispersion does not have any basis in a discrete setting.

The above observations suggest that establishing a rigorous foundation for dispersion in discrete distributions requires an order that retains the properties of the dispersive order while being better suited to the discrete setting. Surprisingly, this gap in the foundation of dispersion measures has neither been explicitly identified nor addressed in the literature. Many important works on the topic, such as \citet{oja}, restrict their attention a priori to continuous distributions with sufficient differentiability properties. In a recent study by \citet{landoEmpDispo}, a modified dispersive order for empirical distributions is defined; however, the remainder of the paper focuses primarily on a similarly modified version of the usual stochastic order. Moreover, as noted after Proposition \ref{thm:dispoRanges}, the issues associated with the dispersive order in discrete distributions are significantly less severe for empirical distributions with the same sample size, which is the primary focus of \citeauthor{landoEmpDispo}'s approach.

Overall, the set of distributions lacking a rigorous foundation for measuring dispersion still includes empirical distributions with differing sample sizes and all lattice distributions. To address this gap in rigor for such a widely used concept, Section \ref{sec:discDisp} is dedicated to establishing a suitable order for these distributions.

\section{The discrete dispersive orders: Definition and properties} \label{sec:discDisp}

\subsection{Definition}
\label{sec:discDispDef}

Let $\Ddists \subset \Rdists$ denote the set of discrete distributions. Since $\Ddists$ contains a number of difficult to handle distributions with virtually no practical use, we limit our considerations to the class of purposive discrete distributions given in the following definition along with a number of subclasses.

\begin{definition}
\label{def:PDdists}
    Let $F \in \Ddists$ be a cdf with probability mass function (pmf) $f$ and let $X \sim F$. The class of {\em purposive discrete distributions} $\PDdists \subseteq \Ddists$ is defined by
    \begin{align*}
        F \in \PDdists \Leftrightarrow& \ \supp(F) \text{ is order-isomorphic to a subset of } \Z \text{ with at least two }\\
        & \, \text{ elements} \\
        \Leftrightarrow & \ \exists \text{ a subset } A \subseteq \Z, |A| \geq 2, \text{and a bijection } \varphi: \supp(F) \to A\\
        & \ \text{such that } x \leq y \Leftrightarrow \varphi(x) \leq \varphi(y) \ \forall x, y \in \supp(F).
    \end{align*}
\end{definition}

Note that there indeed exist non-degenerate discrete distributions in $\Ddists \setminus \PDdists$, i.e.\ discrete distributions that are not order-isomorphic to a subset of the whole numbers. However, the class $\PDdists$ contains all lattice distributions and empirical distributions, which are the most frequently used kinds of discrete distributions.

The following result, which is crucial for the remainder of this work, is only valid for distributions in $\PDdists$. Let $\N = \{1, 2, \ldots\}$.

\begin{proposition}
\label{thm:IS-char}
	Define
	\begin{equation*}
		 \indsetset = \{\Z, \N, -\N\} \cup \{ \{1, \ldots, n\}: n \in \Nz \}
	\end{equation*}
	and
	\begin{align*}
		\idnseqset_A =& \Bigg\{ (x_j, p_j)_{j \in A} \subseteq \R \times (0, 1]: x_i < x_j \text{ for } i < j, p_j > 0 \text{ for } j \in A, \sum_{j \in A} p_j = 1 \Bigg\}\\
        \intertext{for $A \in \indsetset \setminus \Z$ as well as}
		\idnseqset_{\Z} =& \Bigg\{ (x_j, p_j)_{j \in \Z} \subseteq \R \times (0, 1]: x_i < x_j \text{ for } i < j, p_j > 0 \text{ for } j \in \Z, \sum_{j \in \Z} p_j = 1,\\
		& \; \hphantom{\Bigg\{ (x_j, p_j)_{j \in \Z} \subseteq \R \times (0, 1]:} \inf\{j \in \Z: \sum_{i \leq j} p_i \geq \tfrac{1}{2}\} = 0 \Bigg\}.
	\end{align*}
	For any $F \in \PDdists$, there exists a unique index set $A \in \indsetset$ that is order-isomorphic to $\supp(F)$, and there exists a unique sequence $(x_j, p_j)_{j \in A} \in \idnseqset_A$ such that $\Prob(X = x_j) = p_j$ for all $j \in A$.
	This unique association is denoted by $F \cdfis \left( A, (x_j, p_j)_{j \in A} \right)$. $A$ is said to be the {\em indexing set} of $F$ and $(x_j, p_j)_{j \in A}$ is said to be the {\em identifying sequence} of $F$.
\end{proposition}

Note that if $\idnseqset_{\Z}$ were of the same structure as $\idnseqset_A$ for $A \neq \Z$, $F$ could only be uniquely identified up to an arbitrary index shift of the identifying sequence. The additional condition in $\idnseqset_{\Z}$ assures that the index $0$ relates to the median and thereby fixes the sequence.

\begin{sloppypar}
Throughout the remainder of this paper, let $F \cdfis \left( A, (x_j, p_j)_{j \in A} \right)$ and $G \cdfis$ $\left( B, (y_j, q_j)_{j \in B} \right)$. Furthermore, we establish the conventions $x_a = -\infty$ and $F(x_a) = 0$ for $a < \min A$ as well as $x_a = \infty$ and $F(x_a) = 1$ for $a > \max A$, provided that the minimum and the maximum exist, respectively.
\end{sloppypar}

Our proposed discrete dispersive orders consists of two distinct types of comparison. Smaller jump discontinuities are indicative of a more dispersed distribution because the probability mass is split into more and therefore smaller parts. Also, longer constant intervals between the jump discontinuities are also indicative of a more dispersed distribution because the pieces of probabiliy mass are spread out further. One question that remains is which pairs of jump heights and which pairs of constant intervals have to be ordered accordingly for one distribution to be unambiguously more dispersed than the other. To answer this, we define the following relations.
For any indexing set $A\in\indsetset$, we write $\smi{A}=A \setminus \{\min A\}$ and $\sma{A}=A \setminus \{\max A\}$.

\begin{definition}
	\label{def:compj}
	Let $F, G \in \PDdists$.
 \begin{enumerate}
     \item[a)] The relation $\compj{F}{G}$ on the set $A \times B$ is defined by
	\begin{align*}
		  a \stdCompj b &\Leftrightarrow (F(x_{a-1}), F(x_a)) \cap (G(y_{b-1}), G(y_b)) \neq \emptyset.
	\end{align*}
	for $a \in A, b \in B$. The set $R(\compj{F}{G})$ of all $(a, b) \in A \times B$ with $a \compj{F}{G} b$ is said to be the set of $(F, G)$-{\em dispersion-relevant} pairs of indices.\\
	If $F$ and $G$ are fixed, we write $\stdCompj$ instead of $\compj{F}{G}$.

    \item[b)] The relation $\compja{F}{G}$ on the set $\smi{A} \times \smi{B}$ is defined by
		\begin{equation*}
			a \compja{F}{G} b \Leftrightarrow (a \compj{F}{G} b) \land (a-1 \compj{F}{G} b-1)
		\end{equation*}
		for $a \in \smi{A}, b \in \smi{B}$. The set $R(\compja{F}{G})$ of all $(a, b) \in \smi{A} \times \smi{B}$ with $a \compja{F}{G} b$ is said to be the set of $(F, G)$-{\em $\land$-dispersion-relevant} pairs of indices.
		
    \item[c)] The relation $\compjo{F}{G}$ on the set $\smi{A} \times \smi{B}$ is defined by
		\begin{equation*}
			a \compjo{F}{G} b \Leftrightarrow (a \compj{F}{G} b) \lor (a-1 \compj{F}{G} b-1)
		\end{equation*}
		for $a \in \smi{A}, b \in \smi{B}$. The set $R(\compjo{F}{G})$ of all $(a, b) \in \smi{A} \times \smi{B}$ with $a \compjo{F}{G} b$ is said to be the set of $(F, G)$-{\em $\lor$-dispersion-relevant} pairs of indices.
 \end{enumerate}
\end{definition}

We now propose the following two discrete dispersive orders. The reason why we present two proposals is that it is unclear which one is superior, this will be discussed in more detail in Section \ref{sec:crucialProps}.

\begin{definition}
\label{def:disco}
	Let $F, G \in \PDdists$.
	\begin{enumerate}
		\item[a)] $G$ is said to be {\em at least as $\land$-discretely dispersed} as $F$, denoted by $F \discoa G$, if the following two conditions are satisfied:
        \begin{enumerate}
            \item[(i)] $q_b \leq p_a \quad \forall (a, b) \in R(\stdCompj)$,
            \item[(ii)] $x_a - x_{a-1} \leq y_b - y_{b-1} \quad \forall (a, b) \in R(\stdCompja)$.
        \end{enumerate}
		
		\item[b)] $G$ is said to be {\em at least as $\lor$-discretely dispersed} as $F$, denoted by $F \discoo G$, if the following two conditions are satisfied:
        \begin{enumerate}
            \item[(i)] $q_b \leq p_a \quad \forall (a, b) \in R(\stdCompj)$,
            \item[(ii)] $x_a - x_{a-1} \leq y_b - y_{b-1} \quad \forall (a, b) \in R(\stdCompjo)$.
        \end{enumerate}
        \end{enumerate}
\end{definition}

\subsection{Crucial properties}
\label{sec:crucialProps}

The two most crucial properties of a discrete dispersive order are that it coincides with the original dispersive order on the discrete distributions to which the original order can be applied and that it is transitive. While the importance of the first property is obvious, the second property is particularly crucial when using this order to define measures of discrete dispersion, which is our main motivation. A necessary condition for a functional $\tau$ to be considered a discrete dispersion measure is that $X \gendo Y$ implies $\tau(X) \leq \tau(Y)$, where $\gendo$ is a suitable order of discrete dispersion. If there exist three random variables $X, Y, Z$ such that $X \gendo Y \gendo Z$, this implies $\tau(X) \leq \tau(Y) \leq \tau(Z)$. Since `$\leq$' is transitive, this yields $\tau(X) \leq \tau(Z)$. However, if $\gendo$ is not transitive, we get $X \not\gendo Z$, which is very counterintuitive because, according to the measure $\tau$, $X$ and $Z$ are more different with respect to discrete dispersion than $X$ and $Y$ as well as $Y$ and $Z$. The situation gets even worse, if $Z \gendo X$, which contradicts $\tau(X) \leq \tau(Z)$ if any of the inequalities also holds in a strict sense.

Concerning these two crucial properties, we obtain the following results.

\begin{theorem}
\label{thm:DiscoDispoNice}
	If $F, G \in \PDdists$ satisfy $F(D_F) \subseteq G(D_G)$, then the following implication and equivalence hold:
		\begin{equation*}
			F \discoo G \Longrightarrow F \discoa G \Longleftrightarrow F \dispo G.
		\end{equation*}
\end{theorem}

Note that the implication in Theorem \ref{thm:DiscoDispoNice} is strict, i.e.\ that the reverse implication $F \discoa G \Rightarrow F \discoo G$ does not hold in general under the assumption $F(D_F) \subseteq G(D_G)$. A counterexample is obtained by considering $X \sim \mbox{Bin}(1, \frac{1}{2})$ and $\Prob(Y=0) = \frac{1}{2}$, $\Prob(Y=1) = \frac{1}{4}$, $\Prob(Y=\frac{3}{2}) = \frac{1}{4}$.

\begin{theorem}
\label{thm:discoTrans}
	Let $F, G, H \in \PDdists$. The order $\discoo$ is transitive, i.e.\ $F \discoo G$ and $G \discoo H$ implies $F \discoo H$. However, in general, the order $\discoa$ is not transitive.
\end{theorem}

It is worth noting that any dispersion order is in fact only a preorder or quasiorder. This means that they are reflexive and transitive relations, but they are neither antisymmetric nor total. Totality would mean that each pair of distributions is ordered in one of the two possible directions. This does not make sense since the stochastic order is a foundation underneath measures of the same distributional characteristic. Thus, two distributions should only be ordered if the difference between them is beyond any doubt; if the decision is not unambiguous, it is left up to the measure that is built on the foundation of the order. Antisymmetry would mean that the order being valid in both directions would imply that the two distributions are the same. In the case of dispersion orders, this does not make sense since two distributions that only differ by a shift are still considered equivalent with respect to dispersion. Since we have already considered transitivity, this leaves us with reflexivity, which is important since it reinforces that the considered order is weak and not strict in nature, i.e.\ that $\leq$ is the correct symbol as opposed to $<$. It is easy to see that both proposed orders are reflexive.

\begin{proposition}
\label{thm:discoRefl}
	Let $F \in \PDdists$. The orders $\discoa$ and $\discoo$ are both reflexive, i.e.\ $F \discoa F$ and $F \discoo F$.
\end{proposition}

It turns out that both proposed discrete dispersion orders satisfy all except one of the crucial properties. $\discoo$ is transitive but strictly stronger than the original dispersive order on their joint area of applicability. $\discoa$ is equivalent to $\dispo$ on that area but is not transitive. This begs the question of whether there exists a discrete dispersion order that satisfies both properties. While this question cannot be answered rigorously here, a heuristic explanation suggesting that such an order does not exist is given at the end of Section \ref{sec:heuristics}. A notable exception to this problem is given by the class of all lattice distributions.

\begin{proposition}
\phantomsection\label{thm:latticeDisco}
	\begin{enumerate}
		\item[a)] Let $F$ and $G$ be cdf's of lattice distributions with distances $c_F, c_G > 0$ between neighbouring support points. Then, the following statements are equivalent:
        \begin{enumerate}
            \item[(i)] $F \discoa G$,
            \item[(ii)] $F \discoo G$,
            \item[(iii)] $q_b \leq p_a \ \forall (a, b) \in R(\stdCompj)$ and $c_F \leq c_G$.
        \end{enumerate}
		
		\item[b)] The orders $\discoa$ and $\discoo$ are transitive on the set of all lattice distributions.
	\end{enumerate}
\end{proposition}

Proposition \ref{thm:latticeDisco} shows that the limitations of our approach in defining a discrete dispersive order are not relevant for lattice distributions, which is one of the two most important classes of discrete distributions; the other being the class of empirical distributions. First, one does not have to choose one of the two given options for discrete dispersive orders because they coincide. Second, the one remaining discrete dispersive order fulfills all crucial properties stated throughout this subsection.

\subsection{Further properties}

In this subsection, we further legitimize $\discoa$ and $\discoo$ as discrete versions of the dispersive order $\dispo$ by proving that several positive results concerning $\dispo$ are also true for the discrete orders. First, we consider the equivalence classes of the relation $\dispoeq$, which denotes equivalence with respect to the order $\dispo$. Note that $\dispoeq$ is symmetric by definition and inherits the properties of reflexivity and transitivity from $\dispo$. Thus, $\dispoeq$ is an equivalence relation. The equivalence class of any $F \in \PDdists$ with respect to $\dispoeq$ is given by all real shifts of $F$, i.e.\ $\{F(\cdot - \lambda): \lambda \in \R\}$ \citep[see][p.\ 157]{oja}, so distributions that are equivalent with respect to dispersion can only differ in location. The following are the discrete versions of that result.

\begin{theorem}
\label{thm:discoaBothDirections}
	Let $F, G \in \PDdists$. Then, the following three statements are equivalent:
    \begin{enumerate}
        \item[(i)] $F \discoaeq G$,
        \item[(ii)] $F \discooeq G$,
        \item[(iii)] there exists a $\lambda \in \R$ such that $G(t) = F(t - \lambda)$ for all $t \in \R$.
    \end{enumerate}
\end{theorem}

Theorem \ref{thm:discoaBothDirections} particularly states that $F \discoaeq G$ is equivalent to $F \discooeq G$ for all $F, G \in \PDdists$. We can now consider the quotient set of $\PDdists$ by $\discooeq$, denoted by $\PDdists / \discooeq$, and also define both discrete dispersion orders on that set. For all $\mathcal{F}, \mathcal{G} \in \PDdists / \discooeq$, the orders are defined by
\begin{equation*}
	\mathcal{F} \discoo \mathcal{G}, \quad \text{if and only if} \quad \exists F \in \mathcal{F}, G \in \mathcal{G}: F \discoo G,
\end{equation*}
and analogously for $\discoa$. The consideration of these equivalence classes is relevant to the following result.

\begin{proposition}
\label{thm:discoaSuppSubset}
	Let $F, G \in \PDdists$ not belong to the same equivalence class of $\PDdists$ by $\discoaeq$. Then, it follows from $F \discoa G$ that either $\Lm{1}(D_F) = \Lm{1}(D_G) = \infty$ or $\Lm{1}(D_F) < \Lm{1}(D_G)$ holds. The same is true for $\discoo$ in place of $\discoa$.
\end{proposition}

An analogous result also holds for the the original dispersive order. It directly follows from its Definition \ref{def:dispoDilo}a) and from $\Lm{1}(D_F) = \lim_{\alpha \searrow 0} (F^{-1}(1-\alpha) - F^{-1}(\alpha))$, $\Lm{1}(D_G) = \lim_{\alpha \searrow 0} (G^{-1}(1-\alpha) - G^{-1}(\alpha))$.\par
Another result that relates the dispersive order to the supports of the involved distributions is given in \citet[p. 42, Theorem 1.7.6a)]{mueller-stoyan} and can also be reproduced for both discrete dispersive orders.

\begin{proposition}
\label{thm:discoaInfSto}
\begin{sloppypar}
	Let $F, G \in \PDdists$. If $F \discoa G$ and $\min(\supp(F)) \leq$ $\min(\supp(G))$ with both minimums existing, then $F \sto G$. The same is true for $\discoo$ in place of $\discoa$.
\end{sloppypar}
\end{proposition}

Analogously, if $F$ is less dispersed than $G$ with respect to either order, and $\max(\supp(F)) \geq \max(\supp(G))$ holds with both maximums existing, $F \sto G$ also follows.

\subsection{Helpful examples, results and heuristics}
\label{sec:heuristics}

After having defined our two proposals for discrete dispersive orders and having solidified them as such using a number of results, we revisit the motivating real-world Example \ref{exm:para1}.

\begin{example} \label{exm:para2}
\begin{itemize}
\item [(a)]

In this comparison, probability mass transfer arguments pose a very convincing arguments that the distribution of the second sample is unambiguously more dispersed than the first. However, $p \dispo q$ does not hold.

When considering the discrete orders, $p \discoa q$ follows directly from the fact that $p$ and $q$ are lattice distributions with $p_i>q_i$ for $i=0,\ldots,4$. According to Proposition \ref{thm:latticeDisco}, $p \discoo q$ is also true. Thus, both discrete dispersive order behave according to the intuitive perception. Consequently, all the dispersion measures considered in Section \ref{sec:42} that preserve this order are ranked equally.

\item [(b)]
In this comparison, we observe $p_i > q_i$ for $i=0,1,3$ and $p_i = 0$ for $i \geq 4$, but $p_2<q_2$. Thus, the simple probability mass transport argument from part (a) no longer applies and the situation is less obvious.

It follows from Proposition \ref{thm:latticeDisco} that $p\discoa q$. The fact that $p_2<q_2$ does not contradict this because the two jumps from 1 to 2 do not occur at the same probability level in the two cdf's, i.e. $2 \not\stdCompj 2$. This is because the relative frequencies of $0$ and $1$ are much higher in sample 1 compared to sample 2. Thus, $1 \stdCompj 2$, $2 \stdCompj 8$ and $2 \stdCompj 10$, which does not contradict $p\discoa q$ because of $p_1<q_2$, $p_2<q_8$ and $p_2<q_{10}$. (Here, the actual support points were used for the relation $\stdCompj$ instead of the corresponding elements of the indexing sets.)
\end{itemize}
\end{example}

In this subsection, we explain the idea behind the definition of $\discoa$ and $\discoo$ as well as a number of results that clarify how these orders work. First, both orders consist of two different criteria: one comparing the jump heights of two distributions and the other comparing the lengths of their constant intervals.

The condition (i) for the jump heights is the same in the definition of both discrete orders. It stipulates that any jump of the more dispersed cdf needs to be shorter than all jumps of the less dispersed cdf that it overlaps. Intuitively, this means that the probability mass that is contained in one jump of the less dispersed cdf is spread out over more than one jump for the less dispersed cdf. On this basis, the condition can be rewritten as follows.

\begin{proposition}
\label{thm:discoDens}
    Let $F, G \in \PDdists$. Then, the following equivalence holds for condition (i) in Definitions \ref{def:disco}a) and b):
    \begin{equation*}
        q_b \leq p_a \quad \forall (a, b) \in R(\stdCompj) \quad \Leftrightarrow \quad g(G^{-1}(p)) \leq f(F^{-1}(p)) \quad \forall p \in (0, 1),
    \end{equation*}
    where $f, g$ denote the pmf's and $F^{-1}, G^{-1}$ denote the (left-continuous) quantile functions of $F, G$, respectively.
\end{proposition}

This is the discrete analogue of $F \dispo G$ for differentiable cdf's $F$ and $G$ with interval support, which is equivalent to
\begin{equation*}
    g(G^{-1}(p)) \leq f(F^{-1}(p)) \quad \forall p \in (0, 1),
\end{equation*}
where $f$ and $g$ are the Lebesgue densities of $F$ and $G$, respectively.

\begin{example}
\phantomsection\label{exm:2Unif3plus}
	\begin{enumerate}
		\item[a)] This example is visualized in the left panel of Figure \ref{fig:ExmCompj}. The indexing sets $A$ and $B$ of $F$ and $G$ are equal to their supports. The identifying sequences are given by $(j, \frac{1}{2})_{j \in \{1, 2\}}$ and $(j, \frac{1}{5})_{j \in \oneto{5}}$, respectively. We now go through the elements of $A$ one by one, starting with $a=1$, which yields
		\begin{align*}
			a \stdCompj b &\Leftrightarrow (F(x_{a-1}), F(x_a)) \cap (G(y_{b-1}), G(y_b)) \neq \emptyset\\
			&\Leftrightarrow (0, \tfrac{1}{2}) \cap (\tfrac{b-1}{5}, \tfrac{b}{5}) \neq \emptyset\\
			&\Leftrightarrow b \in \{1, 2, 3\}
		\end{align*}
		for $b \in B$. Similarly, for $a=2$, we obtain
		\begin{align*}
			a \stdCompj b &\Leftrightarrow (\tfrac{1}{2}, 1) \cap (\tfrac{b-1}{5}, \tfrac{b}{5}) \neq \emptyset\\
			&\Leftrightarrow b \in \{3, 4, 5\}
		\end{align*}
		for $b \in B$. Since $A = \{1, 2\}$, it follows that
		\begin{equation*}
			R(\stdCompj) = \{(1, 1), (1, 2), (1, 3), (2, 3), (2, 4), (2, 5)\}.
		\end{equation*}
		This is depicted using the cdf's in the left panel of Figure \ref{fig:ExmCompj}, which is very similar to the lower left panel of Figure \ref{fig:CEXdispoDisc}. In fact, the same jumps have the same colour on them, the colouring is simply extended to the entire heights of the jumps in Figure \ref{fig:ExmCompj}. In Figure \ref{fig:CEXdispoDisc}, the colours in the cdf's are associated with the Q-Q-plot in the panel above. This is representative of another characterization of the order $\stdCompj$: $a \stdCompj b$ holds, if and only if the point $(x_a, y_b)$ is part of the corresponding Q-Q-plot. Both formulations mean that the the same piece of probability mass lies on the point $x_a$ in $F$ and on the point $y_b$ in $G$.
		
		\begin{figure}[ht]
			\centering{
				\includegraphics[width=\textwidth]{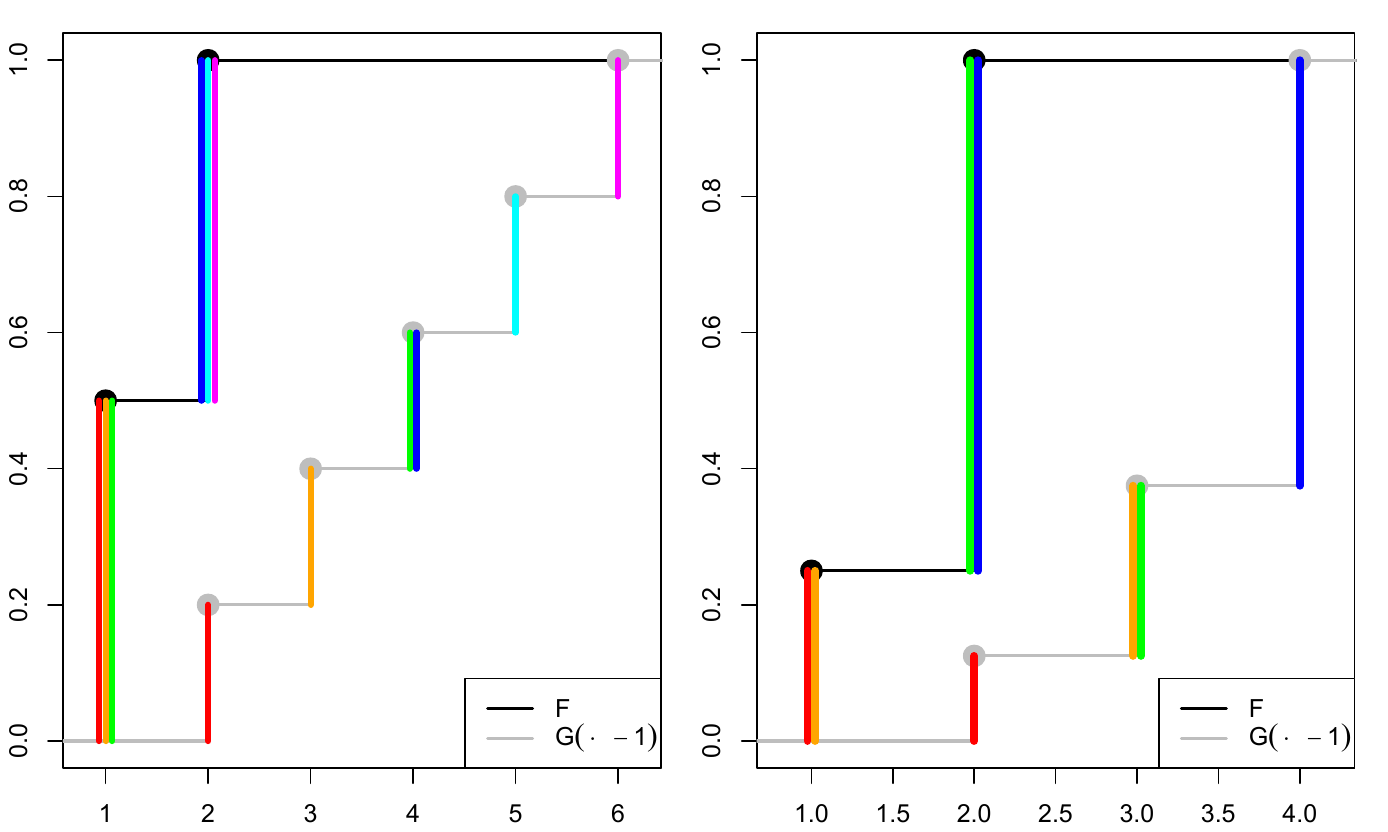}
				\caption{\label{fig:ExmCompj}Visualization of Example \ref{exm:2Unif3plus}a) in the left panel and of Example \ref{exm:2Unif3plus}b) in the right panel. The pairs of jumps, of which the heights are to be compared (and which are therefore connected by the relation $\stdCompj$), are marked with the same colour.}
			}
		\end{figure}
		
		\item[b)] Because of the simple structure of the cdf's in part a), they are not instructive for exploring the connection of the relation $\stdCompj$ to the discrete dispersive order $\discoa$ and $\discoo$. Therefore, we also consider the following pair of cdf's. Let $X \sim F$ and $Y \sim G$ be defined by
		\begin{alignat*}{5}
			\Prob(X = 1) &= \tfrac{1}{4}, \quad &\Prob(X = 2) &=& \; \tfrac{3}{4}&,\\
			\Prob(Y = 1) &= \tfrac{1}{8}, \quad &\Prob(Y = 2) &=& \; \tfrac{1}{4}&, \quad \Prob(Y = 3) = \tfrac{5}{8}.
		\end{alignat*}
		Note that $F$ and $G$ are cdf's of lattice distributions with distance one between neighbouring points in the support; therefore, only condition (i) is relevant in both parts of Definition \ref{def:disco}. Once again, the indexing sets $A$ and $B$ of $F$ and $G$ are given by their supports. By considering whether the jumps of $F$ and $G$ intersect, we obtain
		\begin{equation*}
			R(\stdCompj) = \{ (1, 1), (1, 2), (2, 2), (2, 3) \}
		\end{equation*}
		(see right panel of Figure \ref{fig:ExmCompj}). Going back to the definition of $F$ and $G$, it is obvious that the first jump of $F$ is at least as high as the first two jumps of $G$ and the second jump of $F$ is higher than the last two jumps of $G$, yielding $F \discoa G$ and $F \discoo G$. Note that the third jump of $G$ (height $\frac{5}{8}$) is higher than the first jump of $F$ (height $\frac{1}{4}$), represented by the pair $(1, 3) \in A \times B$ of indices. However, because the jumps do not overlap, $(1, 3) \notin R(\stdCompj)$, i.e.\ the comparison is not relevant to the discrete dispersive orders.
	\end{enumerate}
\end{example}

Unlike in the continuous case, condition (i) in both parts of Definition \ref{def:disco} is not sufficient to unambiguously ensure that $F$ is less dispersed than $G$. It only imposes a requirement on the jump heights of the cdf and not on the distance between the support points. If we chose the support points of $G$ to be very close to each other and the support points of $F$ to be very far from each other, $G$ can not be unambiguously more dispersed than $F$. Interestingly, the corresponding conditions (ii) in both parts of Definition \ref{def:disco} are also directly informed by the behaviour of the original dispersive order $\dispo$. However, while condition (i) is informed by its behaviour on absolutely continuous distributions, both conditions (ii) are informed by its behaviour on discrete distributions.

\begin{proposition}
\label{thm:dispoDisc}
	Let $F, G \in \Ddists$. Then $F \leq_{disp} G$ is equivalent to
	\begin{align*}
		F(D_F) \subseteq G(D_G) \quad \text{and} \quad \Lm{1}(F^{-1}(\{p\})) \leq \Lm{1}(G^{-1}(\{p\})) \ \forall p \in F(D_F),
	\end{align*}
    where $\Lm{1}$ denotes the one-dimensional Lebesgue measure and $F^{-1}(\{p\}), G^{-1}(\{p\})$ denote the preimage of $p$ for the cdf $F, G$, respectively.
\end{proposition}

For the pairs of discrete distributions to which $\dispo$ is applicable, it holds if the constant intervals of the less dispersed cdf are shorter than the constant intervals of the more dispersed cdf at the same function values. Since the image of the less dispersed cdf $F$ is a subset of the image of the more dispersed cdf $G$, it is sufficient to consider the intervals at which both cdf's take values in the image of $F$. Thus, condition (ii) needs to require the constant intervals of $G$ to be longer than the constant intervals of $F$ in some pointwise sense. This makes sense since it overall spreads out the distribution $G$ more than that of $F$. The only question that remains is which constant intervals shall be compared.

For general discrete distributions, we do not have $F(D_F) \subseteq G(D_G)$. Instead, it makes sense to use the already established relation $\stdCompj$ for picking pairs of constant intervals to be compared in a pointwise sense. Since $\stdCompj$ considers whether jumps of discrete cdf's are overlapping, utilizing it for the constant intervals in between the jumps leads to asymmetries. However, these asymmetries cancel each other out if we apply the relation twice, holding either for both jumps on either side of a constant interval or for at least one of them. The first option connects these two conditions using a logical and, giving us condition (ii) in the definition of $\discoa$, whereas the second option connects them using a logical or, giving us condition (ii) of $\discoo$. The two corresponding relations $\stdCompja$ and $\stdCompjo$ also have a much more illustrative interpretation given in the following. Preliminarily, we define the set of (upper and lower) nearest neighbours of a given constant interval.

\begin{definition}
	Let $F, G \in \PDdists$ and let $a \in \smi{A}$. Then, the {\em set of (upper and lower) nearest neighbours of $F$ in $G$ with respect to $a$} (denoted by $\nn{a}{F}{G}$) is defined as follows.
	\begin{enumerate}
		\item[(i)] If $G(D_G) \cap (0, F(x_{a-1})] \neq \emptyset$ and $G(D_G) \cap [F(x_{a-1}), 1) \neq \emptyset$, define
		\begin{equation*}
			\nn{a}{F}{G} = \left\{ \sup\left(G(D_G) \cap (0, F(x_{a-1})]\right), \inf\left(G(D_G) \cap [F(x_{a-1}), 1)\right) \right\}.
		\end{equation*}
		
		\item[(ii)] If $G(D_G) \cap (0, F(x_{a-1})] = \emptyset$ and $G(D_G) \cap [F(x_{a-1}), 1) \neq \emptyset$, define
		\begin{equation*}
			\nn{a}{F}{G} = \left\{ \inf\left(G(D_G) \cap [F(x_{a-1}), 1)\right) \right\}.
		\end{equation*}
		
		\item[(iii)] If $G(D_G) \cap (0, F(x_{a-1})] \neq \emptyset$ and $G(D_G) \cap [F(x_{a-1}), 1) = \emptyset$, define
		\begin{equation*}
			\nn{a}{F}{G} = \left\{ \sup\left(G(D_G) \cap (0, F(x_{a-1})]\right) \right\}.
		\end{equation*}
	\end{enumerate}
	Here, it is impossible that both sets are empty since this would imply $\emptyset = G(D_G) \cap (0, 1) = G(D_G)$ and thus $|\supp(G)| = 1$, which contradicts $G \in \PDdists$. Furthermore, note that $F(x_{a-1})$ is the value that $F$ takes on the interval $[x_{a-1}, x_a)$, which is generally associated with the index $a$.
\end{definition}

\begin{proposition}
\label{thm:discoNN}
	Let $F, G \in \PDdists$ satisfy condition (i) of either part of Definition \ref{def:disco}. Then,
	\begin{enumerate}
		\item[a)] $R(\stdCompja) = \bigcup_{a \in \smi{A}} \left( \{a\} \times \{ \beta \in \smi{B}: G(y_{\beta-1}) \in \nn{a}{F}{G} \} \right)$,
		
		\item[b)] $R(\stdCompjo) = \bigcup_{b \in \smi{B}} \left( \{ \alpha \in \smi{A}: F(x_{\alpha-1}) \in \nn{b}{G}{F} \} \times \{b\} \right)$.
	\end{enumerate}
\end{proposition}

Proposition \ref{thm:discoNN} states that $\stdCompja$ links all constant intervals of $F$ with the nearest neighbouring intervals of $G$ and $\stdCompjo$ links all constant intervals of $G$ with the nearest neighbouring intervals of $F$. Overall, the pairs of indices linked by $\stdCompja$ is a subset of the pairs linked by $\stdCompjo$.

\begin{example}
\phantomsection\label{exm:compjao}
	\begin{enumerate}
		\item[a)] Let $F, G \in \PDdists$. Furthermore, let $A = \{1, 2, 3\}$, $B = \oneto{8}$ and let $p_j = \tfrac{1}{3}$ for all $j \in A$ and
		\begin{equation*}
			(q_1, \ldots, q_8)^\top = \frac{1}{16} \left(4, 1, 1, 2, 2, 1, 1, 4\right)^\top.
		\end{equation*}
        Using Proposition \ref{thm:discoNN}, we have
		\begin{align*}
			R(\stdCompja) &= \{ (2, 3), (2, 4), (3, 6), (3, 7) \},\\
			R(\stdCompjo) &= \{ (2, 2), (2, 3), (2, 4), (3, 4), (2, 5), (3, 5), (2, 6), (3, 6), (3, 7), (3, 8) \}.
		\end{align*}
		
		\begin{figure}[t]
			\centering{
				\includegraphics[width=\textwidth]{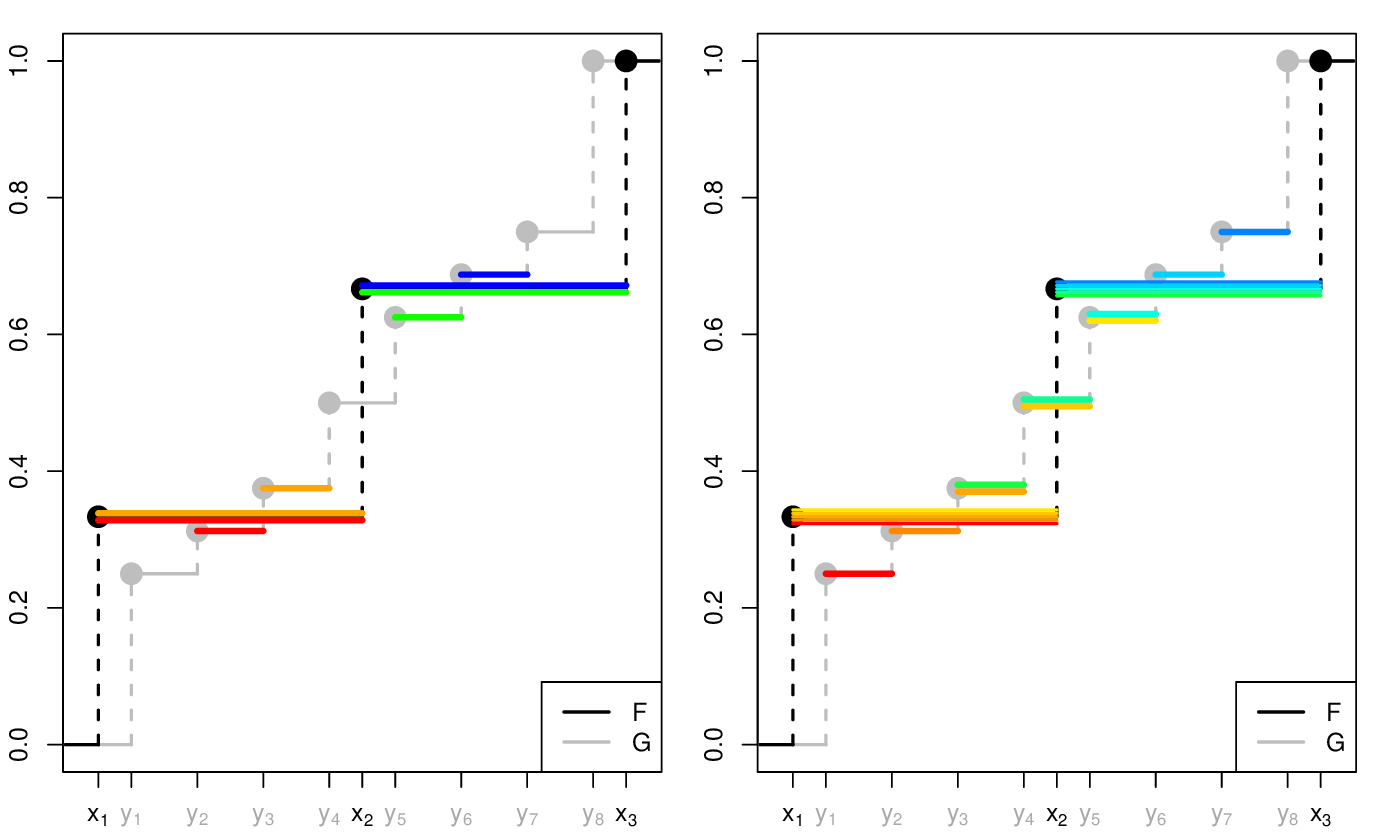}
				\caption{\label{fig:ExmCompjao}Visualization of Example \ref{exm:compjao}. The pairs of constant intervals, of which the lengths are to be compared with respect to $\stdCompja$ (in the left panel) and $\stdCompjo$ (in the right panel), are marked with the same colour.}
			}
		\end{figure}
		
		Since $G$ as the more dispersed cdf has more constant intervals than $F$ (since the jump heights of $G$ are shorter), the number of comparisons for $\discoo$ in the right panel is a lot higher. For $\discoa$ in the left panel, some constant intervals of $F$ are not compared to any intervals of $G$. However, because of the higher number of intervals of $G$, the smaller number of comparisons made by $\discoa$ seems sufficient to ensure that $G$ is unambiguously more dispersed than $G$. Unfortunately, the smaller amount of comparisons implies $\discoa$ is not transitive.
	\end{enumerate}
\end{example}

This additional information on $\discoa$ and $\discoo$ intuitively explains Proposition \ref{thm:latticeDisco} because only condition (i) is active for both orders as the distance between support points is fixed by only considering lattice distributions. Therefore, both discrete orders are equivalent and behave nicely. However, if the orders are applied to non-lattice distributions, Theorems \ref{thm:DiscoDispoNice} and \ref{thm:discoTrans} state that both orders lack one crucial property. In the following, we heuristically explain why transitivity and equivalence to $\dispo$ in the case $F(D_F) \subseteq G(D_G)$ seem to contradict each other when comparing the lengths of constant intervals of discrete cdf's.

\begin{figure}[ht]
	\centering{
		\includegraphics[width=\textwidth]{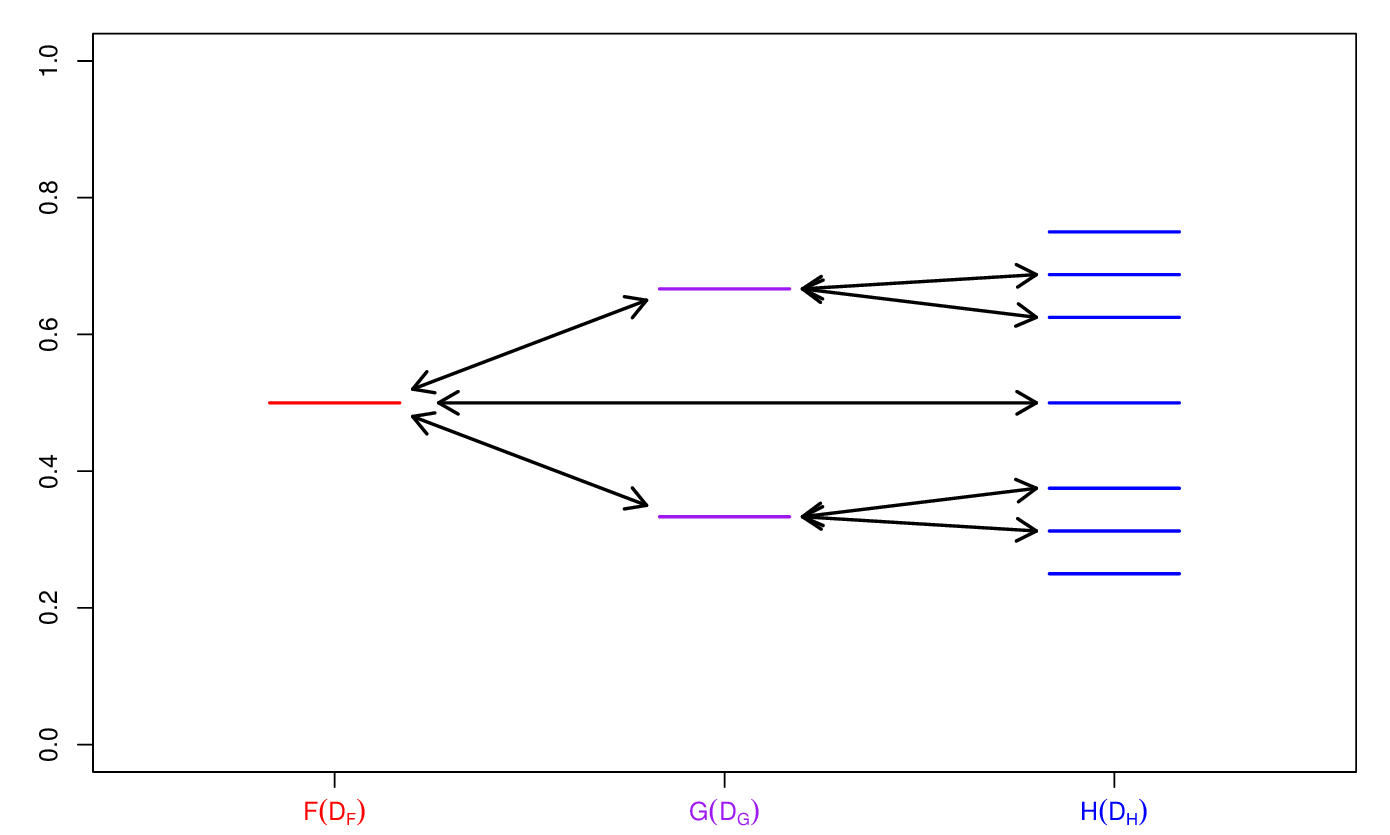}
		\caption{\label{fig:CounterexmTransDiscoa}Illustration of the heuristic explanation why a discrete dispersive order cannot be both equivalent to $\dispo$ in the case $F(D_F) \subseteq G(D_G)$ and transitive. Each horizontal line symbolizes an element of $F(D_F)$, $G(D_G)$ and $H(D_H)$, respectively, and therefore also the constant intervals of the corresponding cdf's. The linked pairs of constant intervals are identified by double-sided arrows. Each constant interval is represented by the value that the corresponding cdf takes on there.}
	}
\end{figure}

If we consider the exemplary situation depicted in Figure \ref{fig:CounterexmTransDiscoa}, we directly see that $F(D_F) \subseteq H(D_H)$ holds. Thus, a necessary condition for $F$ being unambiguously less dispersed than $H$ is that the two corresponding constant intervals need to be linked in the sense that interval of $H$ needs to be longer than that of $F$. Otherwise, the corresponding order could not be equivalent to $\dispo$. However, since neither $F(D_F) \subseteq G(D_G)$ nor $G(D_G) \subseteq H(D_H)$ holds, the constant intervals of $F$ are compared with the nearest neighbours in $G$ and the same is done from $G$ to $H$. Having the more dispersed cdf dictate the comparisons instead of the less dispersed cdf would again contradict the equivalence to $\dispo$. However, by linking the one interval of $F$ to intervals of $H$ via $G$ then leads to vastly different links than connecting the one interval of $F$ directly to the intervals of $H$. This means that the comparisons are not transitive, and by choosing the lengths of the constant intervals suitably, neither is a corresponding order of dispersion.

\section{Discrete dispersion measures} \label{sec:discDispMeas}

In this section, we consider a number of well-known measures of dispersion in the classical sense of Definition \ref{def:dispMeas} and analyze their compatibility with the discrete dispersion orders derived in Section \ref{sec:discDisp}. If a measure preserves these orders, it reliably and meaningfully measures not only the dispersion of continuous distributions, but also of discrete distributions. However, if a measure does not preserve either discrete dispersive order, this presents a rigorous argument for discouraging the use of this dispersion measure for discrete distributions.

\subsection{Relationships to other orders of dispersion} \label{sec:otherDispOrders}

Analyzing the relationship to other dispersion orders is not only helpful in terms of integrating the discrete dispersive orders into an existing framework, but also simplifies the proof of whether certain dispersion measures preserve the discrete dispersive orders. For example, since the standard deviation is centered around the mean, just like the dilation order, it is much easier to show that the standard deviation preserves the dilation order than to show the analogous statement directly for the dispersive order.

Before coming to the dilation order as the most well-known alternative dispersion order toward the end of this subsection, we start out with another order of dispersion that is related to the usual stochastic order. The so-called weak dispersive order was introduced and noted to be weaker than $\dispo$ by \citet[p.\ 326]{weakDispOrder}, who used it as a starting point for a multivariate dispersion order. They said that $F$ precedes $G$ in the weak dispersive order, if $|X - X'| \sto |Y - Y'|$ holds for $X, X' \sim F$ independent and $Y, Y' \sim G$ independent.

\begin{theorem}
\label{thm:discoaWD}
	Let $F, G \in \PDdists$ with $X, X' \sim F$ independent and $Y, Y' \sim G$ independent. Then, $F \discoa G$ implies $|X - X'| \sto |Y - Y'|$.
\end{theorem}

Obviously, it follows directly that $\discoo$ is also stronger than the weak dispersive order.

In order to see that the reverse implication of Theorem \ref{thm:discoaWD} is not true, we make use of the fact that $\discoa$ is not transitive. For that, let $F, G, H \in \PDdists$ be defined as in the part of the proof of Theorem \ref{thm:discoTrans}b), where the counterexample for the transitivity of $\discoa$ was constructed. It is shown there that $F \discoa G$ and $G \discoa H$, but $F \not\discoa H$. If we now let $X, X' \sim F$ independent, $Y, Y' \sim G$ independent and $Z, Z' \sim H$ independent, Theorem \ref{thm:discoaWD} yields $|X-X'| \sto |Y-Y'|$ as well as $|Y-Y'| \sto |Z-Z'|$. Since the stochastic order is transitive, we have now shown $|X-X'| \sto |Z-Z'|$ while $F \not\discoa H$ holds and, thus, that the statement of Theorem \ref{thm:discoaWD} is indeed a strict implication.

\bigskip

We now turn our attention to the dilation order $\dilo$, which is a weakening of the original dispersive order $\dispo$ (see Proposition \ref{thm:diloProps}b)). The proof of this implication given by \citet[pp.\ 158--159]{oja} employs a third order as an intermediate step. This intermediate order is denoted by $\leq_1^*$ by \citeauthor{oja} and $F \leq_1^* G$ is said to hold, if, under the assumption of equal means, the cdf's $F$ and $G$ intersect exactly once with $F$ being smaller than $G$ before the intersection and larger afterwards.

The fact that the discrete dispersive orders also imply the dilation order is proved in a similar way. However, the intersection criterion in this case is not as simple as for $\dispo$. The following lemma gives the corresponding result, which is the discrete analogue of $\leq_1^*$.

\begin{lemma}
\label{thm:discoaIntersecCrit}
	Let $F, G \in \PDdists$ with $F \neq G$ have finite and coinciding means and satisfy $F \discoa G$. Then:
	\begin{enumerate}
		\item[a)] $\exists (a, b) \in A \times \smi{B}: F(x_{a-1}) \leq G(y_{b-1}) \leq F(x_a), y_{b-1} < x_a \leq y_b$.
		\item[b)] One of the following two statements is true:
		\begin{enumerate}
			\item[(i)] $\exists (a, b) \in \sma{A} \times \sma{B}: F(x_{a}) = G(y_{b}), y_b < x_a, x_{a+1} \leq y_{b+1}$ \qquad or
			\item[(ii)] $\exists (a, b) \in A \times \smi{B}: F(x_{a-1}) < G(y_{b-1}) < F(x_a), y_{b-1} < x_a \leq y_b$.
		\end{enumerate}
	\end{enumerate}
\end{lemma}

\begin{figure}
	\centering{
		\includegraphics[width=\textwidth]{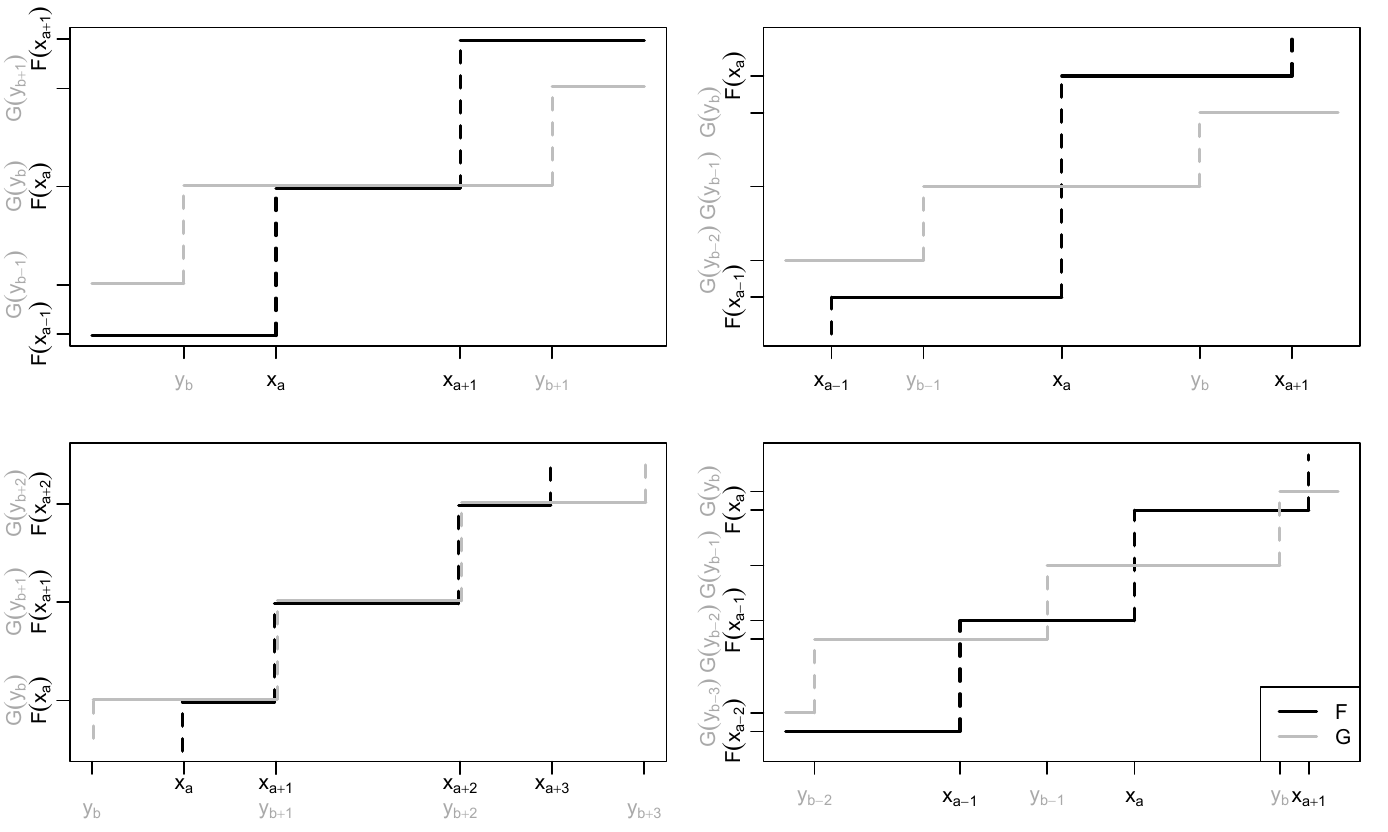}
		\caption{\label{fig:discoaIntersecCrit}Different variations of 'generalized intersections' between two standardized (w.r.t. the mean) cdf's $F$ and $G$ with $F \discoa G$, as specified in Lemma \ref{thm:discoaIntersecCrit}b). Left panels: condition (i). Right panels: condition (ii).}
	}
\end{figure}

As mentioned before, the statement of Lemma \ref{thm:discoaIntersecCrit} is the discrete analogue of an intersection of $F$ and $G$ in the continuous case, which is the transition from $G$ being larger to $F$ being larger. Note that part b) is just a more refined version of part a) and distinguishes between two kinds of intersection equivalents, both of which are depicted schematically in Figure \ref{fig:discoaIntersecCrit}.

If condition (i) is fulfilled, the images of the two (standardized) cdf's share a common element and the generalized intersection occurs as both cdf take that value. This means that (the more dispersed cdf) $G$ is larger than $F$ before said constant interval. After the constant interval, $G$ is either smaller than $F$ right away (as in the upper left panel of Figure \ref{fig:discoaIntersecCrit}) or the two cdf's coincide for a while before $G$ eventually becomes smaller (as in the lower left panel). The specific formulation of condition (i) (and also condition (ii)) that allows equality on the right side but not on the left is somewhat arbitrary in the sense that one could swap the requirements without invalidating the result. Although this potentially results in a different pair $(a, b)$ indicating at which point the intersection takes place, that pair could be used in the same way going forward. Note that if equality was disallowed on both sides, Lemma \ref{thm:discoaIntersecCrit} would no longer be true.\par
If condition (ii) is fulfilled, a jump of $F$ and a constant interval of $G$ form a cross (or a kind of degenerated cross if equality holds on the right side as discussed in the previous paragraph). $G$ is either larger than $F$ before that cross and smaller after (as exemplified in the upper right panel of Figure \ref{fig:discoaIntersecCrit}) or this kind of situation can occur repeatedly (as exemplified in the lower right panel). The latter situation is the main difficulty in the proof of the following theorem.

\begin{theorem}
\label{thm:discoaDilo}
	Let $F, G \in \PDdists$ have finite means. Then, $F \discoa G$ implies $F \dilo G$.
\end{theorem}

According to Theorems \ref{thm:discoaWD} and \ref{thm:discoaDilo} and, to a lesser extent, Lemma \ref{thm:discoaIntersecCrit}, the discrete dispersive orders relate to other orders of dispersion for discrete distributions in the same way as the original dispersive order does for continuous distributions.

\subsection{Compatibility of the discrete orders with popular dispersion measures} \label{sec:42}

With the additional insights from Section \ref{sec:otherDispOrders}, it is now much easier to see which well-known continuous dispersion measures also qualify as discrete dispersion measures. The considered dispersion measures are introduced along with refrences for their fulfilment of properties (D1) and (D2) from Definition \ref{def:dispMeas}. Most of the references only prove (D2) since (D1) is obtained easily by utilising basic properties of the mean, quantiles and expectiles.

\begin{proposition}
\label{thm:popDispMeas}
	For $F \in \Rdists$, let $X, X' \sim F$ be independent and let $e_X: (0, 1) \to \R$ be the corresponding expectile function in the case $F \in \Intdists{1}$.
	\begin{enumerate}
		\item[a)] The {\em standard deviation}
		\begin{equation*}
			\sd: \Intdists{2} \to [0, \infty), \quad F \mapsto \sqrt{\E[(X - \E[X])^2]}
		\end{equation*}
		satisfies conditions (D1) and (D2), see \citet[p.\ 159]{oja}.
		
		\item[b)] The {\em Gini mean difference}
		\begin{equation*}
		\phantomsection\label{eqn:gmd}
			\gmd: \Intdists{1} \to [0, \infty), \quad F \mapsto \E[|X - X'|]
		\end{equation*}
		satisfies conditions (D1) and (D2), see \citet[p.\ 160]{oja}.
		
		\item[c)] The {\em mean absolute deviation from the mean}
		\begin{equation*}
			\mad: \Intdists{1} \to [0, \infty), \quad F \mapsto \E[|X - \E[X]|]
		\end{equation*}
		satisfies conditions (D1) and (D2), see \citet[p.\ 15]{hurli}.
		
		\item[d)] The {\em mean absolute deviation from the median}
		\begin{equation*}
			\mdmad: \Intdists{1} \to [0, \infty), \quad F \mapsto \E[|X - F^{-1}(\tfrac{1}{2})|]
		\end{equation*}
		satisfies conditions (D1) and (D2), see \citet[p.\ 15]{hurli}.
		
		\item[e)] For $0 < \alpha < \beta < 1$, the {\em $(\alpha, \beta)$-interquantile range}
		\begin{equation*}
			\iqnr{\alpha}{\beta}: \Rdists \to [0, \infty), \quad F \mapsto F^{-1}(\beta) - F^{-1}(\alpha)
		\end{equation*}
		satisfies conditions (D1) and (D2) by definition.
		
		\item[f)] For $0 < \alpha < \tfrac{1}{2} < \beta < 1$, the {\em $(\alpha, \beta)$-interexpectile range}
		\begin{equation*}
			\ienr{\alpha}{\beta}: \Intdists{1} \to [0, \infty), \quad F \mapsto e_X(\beta) - e_X(\alpha)
		\end{equation*}
		satisfies conditions (D1) and (D2), see \citet[p.\ 517]{ek-exlDisp}.
	\end{enumerate}
\end{proposition}

The measures $\iqnr{\alpha}{\beta}$ and $\mdmad$ involving quantiles require minor assumptions to satisfy (D1) in a discrete setting. It is sufficient to assume that the involved cdf is strictly increasing in the points at which the quantile function is evaluated. These additional assumptions are not necessary, if the alternative definition of the quantile $F^{-1}(p), p \in (0, 1),$ given by
\begin{equation*}
	\tfrac{1}{2} \big( \inf \{t \in \R: F(t) \geq p\} + \sup \{t \in \R: F(t) \leq p\} \big)
\end{equation*}
is utilized, which is often done for empirical quantiles.

The fact that most of the measures from Definition \ref{thm:popDispMeas} preserve $\discoa$ follows directly from Theorems \ref{thm:discoaWD} and \ref{thm:discoaDilo}. Obviously, if a measure preserves $\discoa$, it also preserves $\discoo$.

\begin{corollary}
    The mappings $\sd$, $\mad$ and $\gmd$ all preserve the order $\discoa$.
\end{corollary}

For $\sd$ and $\mad$, this follows from Theorem \ref{thm:discoaDilo} since both $t \mapsto t^2$ and $t \mapsto |t|$ are convex functions on the real numbers. For $\gmd$, it follows from Theorem \ref{thm:discoaWD} since the expected value is a measure of (central) location, which preserves the usual stochastic order $\sto$. We can also use Theorem \ref{thm:discoaDilo} to show that $\gmd$ preserves $\discoa$ since $F \dilo G$ implies $\gmd(F) \leq \gmd(G)$, as proved in \citet[p.\ 126, Thm.\ 2.2]{gmdDiloOld} and pointed out by \citet[p.\ 65]{gmdDiloNew}. Conversely, the fact that $\sd$ preserves $\discoa$ also follows from Theorem \ref{thm:discoaWD} because of
\begin{align*}
	\sd(F)^2 &= \E[X^2] - \E[X]^2 
    = \tfrac{1}{2} \left( \E[X^2] - 2 \E[X] \E[X'] + \E[{X'}^2] \right) \\
    &= \tfrac{1}{2} \E[(X - X')^2]
\end{align*}
for $X, X' \sim F$ independent.\par
It can be shown in a similar way that $\ienr{\alpha}{\beta}$ preserves the order $\discoa$ for $0 < \alpha < \frac{1}{2} < \beta < 1$, which includes all cases relevant for applications. This again holds due to Theorem \ref{thm:discoaDilo}, combined with the fact that $F \dilo G$ implies $\ienr{\alpha}{\beta}(F) \leq \ienr{\alpha}{\beta}(G)$ for all $0 < \alpha < \frac{1}{2} < \beta < 1$. This implication was first shown by \citet[p.\ 2020, Thm.\ 3(b)]{bellini}; a more elementary proof that also includes the reverse implication is given by \citet[p.\ 517]{ek-exlDisp}. The assumptions can be weakened to include all distributions in $\PDdists \cap \Intdists{1}$ without changing the proof.

\begin{corollary}
	If $0 < \alpha < \frac{1}{2} < \beta < 1$, the mapping $\ienr{\alpha}{\beta}$ preserves the order $\discoa$.
\end{corollary}

It remains to be determined whether the mappings $\iqnr{\alpha}{\beta}$ and $\mdmad$ preserve $\discoa$. Both mappings are based on quantiles, which are well-known to be not as useful for discrete as for continuous distributions. This is partly due to the fact that they are not unique in the discrete case. Furthermore, quantiles only evaluate a distribution in a very local sense, which also explains their popularity in robust statistics. However, for discrete distributions, where the probability mass is very sparse, this leads to a lack of information that is conveyed by single evaluations of quantile functions. In accordance with these observations, the interquantile range $\iqnr{\alpha}{\beta}$ does generally not preserve the discrete dispersive orders, as noted in the following result.

\begin{theorem}
\label{thm:CexmDiscoaIQR}
	For all choices $0 < \alpha < \beta < 1$, the mapping $\iqnr{\alpha}{\beta}$ does not preserve the order $\discoa$.
\end{theorem}

\begin{figure}
	\centering{
		\includegraphics[width=\textwidth]{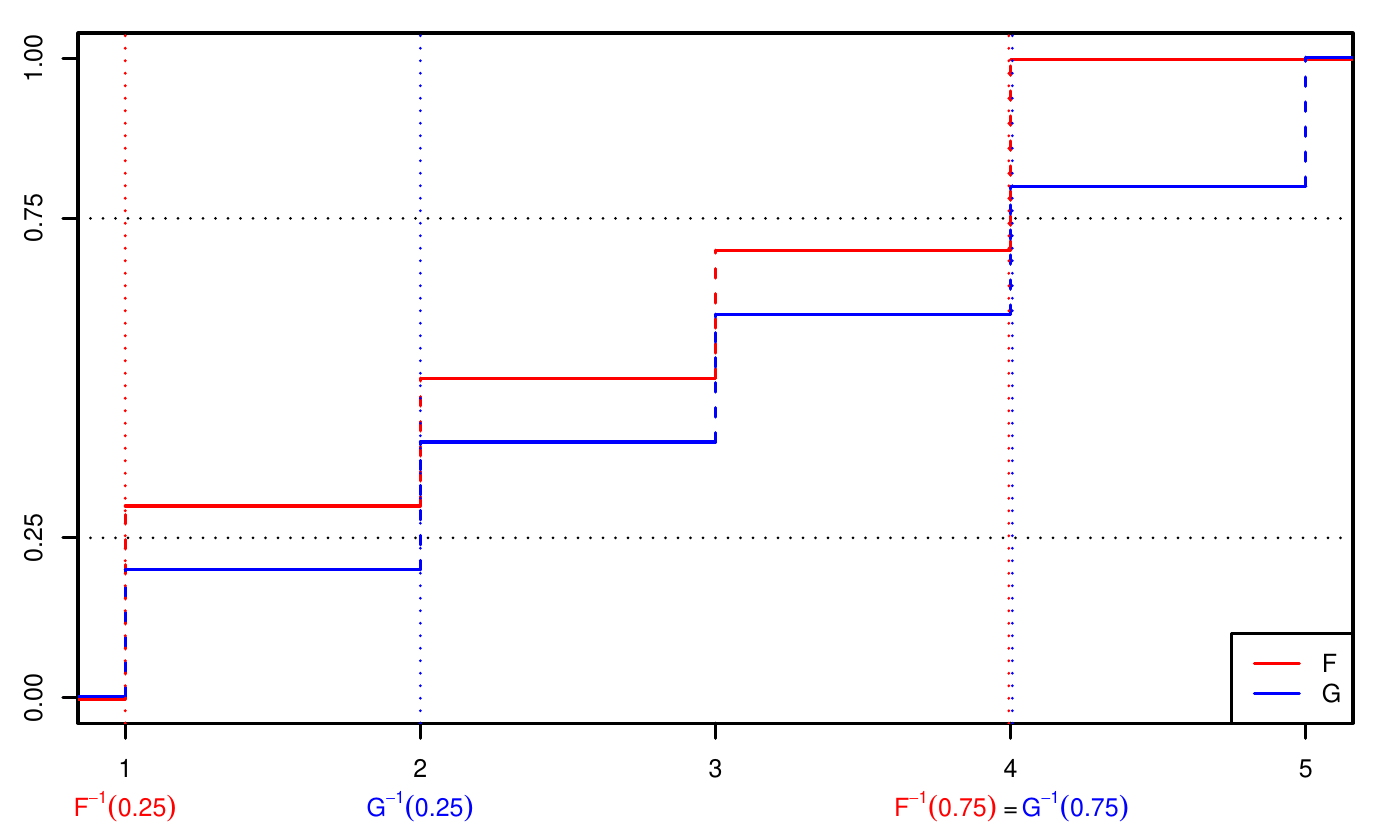}
		\caption{\label{fig:CexmDiscoaIQR}Illustration of a counterexample for Theorem \ref{thm:CexmDiscoaIQR} with $\alpha = \frac{1}{4}$, $\beta = \frac{3}{4}$.}
	}
\end{figure}

The statement of Theorem \ref{thm:CexmDiscoaIQR} also holds, if we replace $\discoa$ by $\discoo$. This is due to the fact that the distributions used in the proof of Theorem \ref{thm:CexmDiscoaIQR} are lattice distributions, for which both discrete dispersive orders are equivalent (see Proposition \ref{thm:latticeDisco}a)).

A counterexample for the interquartile range $\iqnr{\frac{1}{4}}{\frac{3}{4}}$ is depicted in Figure \ref{fig:CexmDiscoaIQR}. The cdf $F$ is specified by by $\Prob(X = 1) = \Prob(X = 4) = \frac{3}{10}$ and $\Prob(X = 2) = \Prob(X = 3) = \frac{1}{5}$ while $G$ is specified by $Y \sim \Unif(\{1, \ldots, 5\})$. Heuristically, to obtain $G$, a third of the probability mass on the outer two jumps of $F$ is shifted outwards and the two pieces are combined to obtain an additional jump. In accordance with this heuristic explanation, it is easy to show that $F \discoa G$ holds since all jump heights of $G$ are smaller than all jump heights of $F$. In particular, $F$ is strictly less dispersed because its support is a subset of the support of $G$. However, Figure \ref{fig:CexmDiscoaIQR} clearly shows that the interquartile range of $G$ is smaller than that of $F$, meaning that the measure is obviously misrepresenting their relationship with respect to dispersion. In applied sciences, where dispersion measures are trusted to be meaningful, this could lead to severe misinterpretations of the data.

Only one of the two drawbacks of using quantile-based measures on discrete distributions is actually relevant for the failure of the interquantile range. Just like in the counterexample depicted in Figure \ref{fig:CexmDiscoaIQR}, all quantiles used for the proof of Theorem \ref{thm:CexmDiscoaIQR} are unique. Thus, the result is also valid for all alternative definitions of quantiles. The failure of the interquantile range is merely due to the local evaluation of the involved distributions, which is not suitable for the sparse nature of discrete distributions. This incompatibility is particularly serious for distributions with small supports.

Theorem \ref{thm:CexmDiscoaIQR} implies that the interquantile range is not fit to be used as a dispersion measure for discrete distributions. This statement is similar to \citet[p.\ 1852]{IER} suggesting that the interquantile range is not a `true measure of variability' because it does not preserve the dilation order. However, it is obvious from the definition of the original dispersive order $\dispo$ that the interquantile range indeed measures dispersion in a meaningful way for continuous distributions. The requirement that dispersion measures should preserve the dilation order is simply too strong. Neither the interquantile range nor the mean absolute deviation from the median could be proved to meet this requirement.

The fact that $\mdmad$ preserves the order $\discoa$ cannot be established as a simple corollary as for all other positive results in this section. In particular, $F \discoa G$ does not generally imply $|X - F^{-1}(\frac{1}{2})| \sto |Y - G^{-1}(\frac{1}{2})|$; a counterexample is given by
\begin{alignat*}{7}
	\Prob(X = 0) &=& \frac{1}{2},\ \Prob(X = 1) &= \Prob(X = 2) \; &=& \; \frac{1}{4},\\
	\Prob(Y = 0) &=& \frac{3}{8},\ \Prob(Y = 1) &= \Prob(Y = 2) \; &=& \; \frac{1}{4},\ \Prob(Y = 3) = \frac{1}{8}.
\end{alignat*}
However, the implication still holds, as shown in the following.

\begin{theorem}
\label{thm:discoaMdmad}
	The mapping $\mdmad$ preserves the order $\discoa$.
\end{theorem}

We conclude this section on discrete dispersion measures by revisiting our real-world examples discussed in \ref{exm:para1} and \ref{exm:para2}.

\begin{example} \label{exm:para3}
\begin{itemize}
\item [(a)]

The obvious difference in dispersion between the two samples is represented by $p \discoa q$ and $p \discoo q$. Consequently, all the dispersion measures considered in Section \ref{sec:42} that preserve this order are ranked equally.
Specifically, the standard deviations of the two samples are $\sd_1=1.29$ and $\sd_2=7.8$; the mean absolute deviations around the mean are $\mad_1=1.03$ and $\mad_2=4.45$; and the Gini mean differences are $\gmd_1=1.33$ and $\gmd_2=5.98$. The large differences in these measures emphasize the clear difference in dispersion, which is also represented in the interquartile ranges $\iqnr{0.25}{0.75}_1 = 2$ and $\iqnr{0.25}{0.75}_1 = 5$.

\item [(b)]
The difference in dispersion here is less obvious than in part (a), but still supported by both discrete dispersive orders. Thus, dispersion measures such as the standard deviation must be smaller for the first sample than for the second. Indeed, the standard deviations of the two samples are $\sd_1=0.76$ and $\sd_2=2.73$; the mean absolute deviations around the mean are $\mad_1=0.55, \mad_2=1.76$; and the Gini mean differences are $\gmd_1=0.61, \gmd_2=2.18$. The smaller differences compared to part (a) are representative of the less clear-cut comparison. The corresponding interquartile ranges $\iqnr{0.25}{0.75}_1 = 1 = \iqnr{0.25}{0.75}_2$ are the same, which does not pose a counterexample proving Theorem \ref{thm:CexmDiscoaIQR}, but still further documents the counterintuitive behaviour of this measure for discrete distributions.
\end{itemize}
\end{example}

\section{Behaviour of discrete distributions in the new framework}
\label{sec:specDists}

In this section, we analyze whether popular families of discrete distributions preserve the discrete dispersive orders. Since all of the distributions considered in the following are lattice distributions with defining distance equal to one, both previously defined discrete dispersive orders are equivalent and it is sufficient to consider condition (i) in the definitions of the discrete orders (see Proposition \ref{thm:latticeDisco}a)). In the formulation of the results, the order $\discoa$ is used since it is the discrete order that is closest to $\dispo$. The only considered non-lattice distribution is the discrete uniform distribution on arbitrary finite sets, which is discussed at the end of the following subsection.

The analysis of the binomial distribution and the Poisson distribution is deferred to the supplement.

\subsection{Discrete uniform and empirical distribution}

The discrete uniform distribution is the simplest discrete distribution. In its usual variant, it puts the same amount of probability mass on a finite number of points that are equidistantly spaced with distance $1$. Since all of the dispersion orders and measures considered in the previous chapters are location invariant, it is sufficient to consider uniform distributions with supports $\{1, \ldots, n\}, n \in \Nz$. If $\Prob(X = k) = \frac{1}{n}$ for all $k \in \{1, \ldots, n\}$, we denote this by $X \sim \Unif[n]$. These distributions are also used in Example \ref{exm:2Unif1} in order to establish that the original dispersive order is far from sufficient for discrete distributions. However, it is easy to show that any two discrete uniform distributions are ordered with respect to the discrete dispersive orders introduced in this work.

\begin{proposition}
	Let $n, m \in \Nz, n < m,$ and let $X \sim \Unif[n]$ and $Y \sim \Unif[m]$. Then, $X \discoa Y$ holds.
	
\end{proposition}

\begin{figure}
	\centering{
		\includegraphics[width=\textwidth]{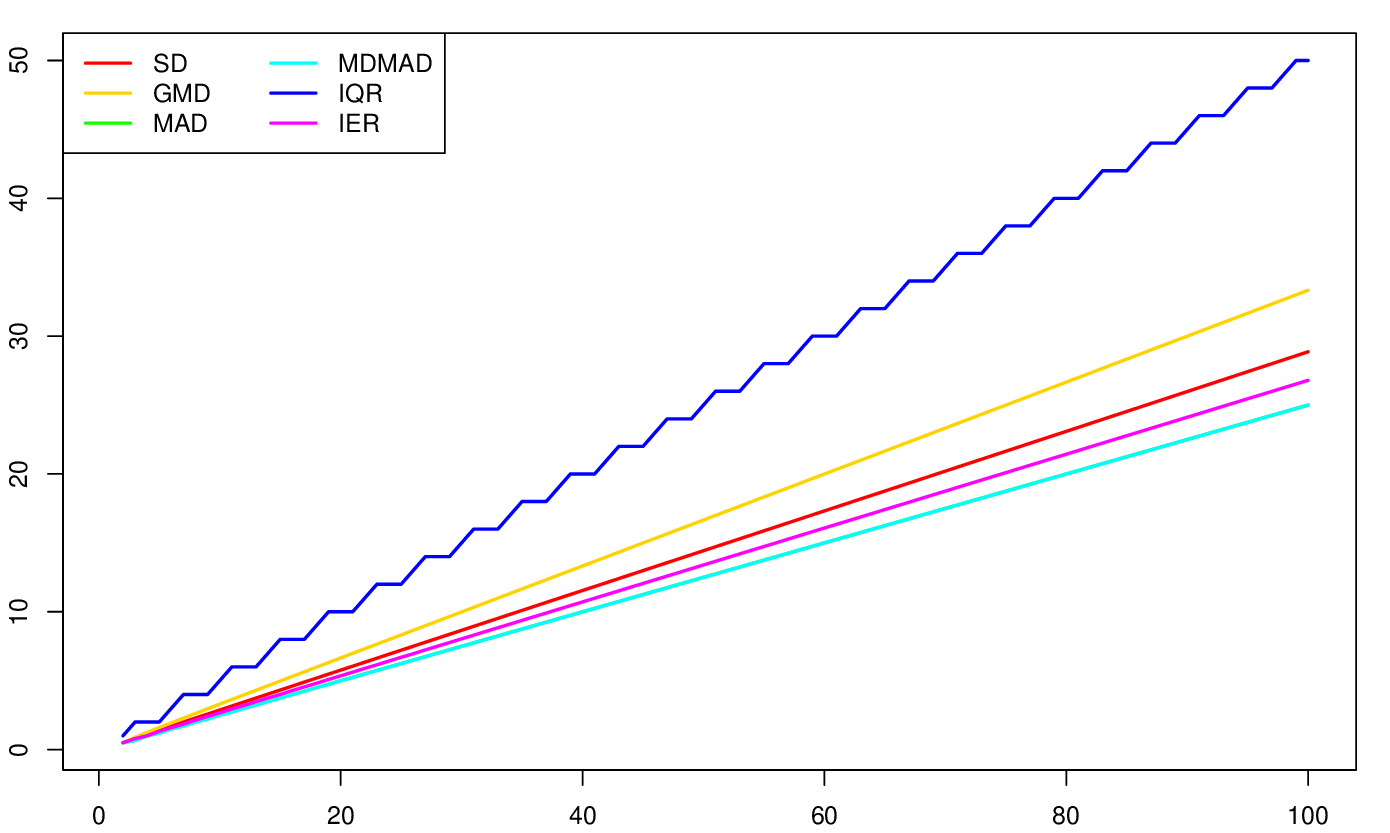}
		\caption{\label{fig:DispMeasDunif}Plot of $\tau(X)$ for six different dispersion measures $\tau$ and $X \sim \Unif[n]$, as a function of $n \in \{2, \ldots, 100\}$. The measures $\mad$ and $\mdmad$ coincide here because of symmetry.}
	}
\end{figure}

The behaviour of the dispersion measures from Section \ref{sec:discDispMeas} for discrete uniform distributions as a function of the parameter $n$ is depicted in Figure \ref{fig:DispMeasDunif}. It shows that five of the six dispersion measures are almost linearly increasing as a function of $n$, although the slight deviations from linearity can barely be seen in Figure \ref{fig:DispMeasDunif}. The average slopes differ between the measures; only $\mad$ and $\mdmad$ are exactly the same since the distribution is symmetric. The graph of the interquartile range $\iqr$ has a different shape since it only takes values in the natural numbers when applied to a lattice distribution with defining distance $1$. This lack of granularity is also somewhat indicative of its lack of compatibility with discrete distributions that is formalized in Theorem \ref{thm:CexmDiscoaIQR}. However, a counterexample for the proof of Theorem \ref{thm:CexmDiscoaIQR} cannot be constructed using this class of discrete uniform distributions.\par
The concept of discrete uniform distributions can be generalized to arbitrary finite sets. Let $S \subset \R$ with $|S| = n \in \Nz$. If now $\Prob(X = s) = \frac{1}{n}$ for any $s \in S$, then $X$ is discretely uniformly distributed on $S$, denoted by $X \sim \Unif(S)$. Note that the set of all generalized discrete uniform distributions is equal to the set of all non-tied empirical distributions. Because of the complexity of this family of distributions, we refrain from trying to obtain general results with respect to discrete dispersive orders. Instead, we discuss a number of special cases for the order $\discoa$ in the following example.

\begin{example}
	Let $S, T \subset \R$ with $2 \leq |S| = m < n = |T|$ and let $X \sim \Unif(S)$, $Y \sim \Unif(T)$.
	\begin{enumerate}
		\item[a)] Let $n$ be a multiple of $m$, so there exists a $c \in \Nz$ such that $n = c \cdot m$. Because of 
		\begin{align*}
			F(D_F) &= \{\tfrac{k}{m}: k \in \oneto{m-1}\} = \{\tfrac{c \cdot k}{n}: k \in \oneto{m-1}\}\\
			&\subset \{\tfrac{k}{n}: k \in \oneto{n-1}\} = G(D_G),
		\end{align*}
		$F \discoa G$ is equivalent to $F \dispo G$ in this case. According to Proposition \ref{thm:dispoDisc}, $F \discoa G$ holds if $x_{k+1} - x_k \leq y_{c \cdot k + 1} - y_{c \cdot k}$ holds for all $k \in \oneto{m-1}$. Hence, $m-1$ comparisons are made overall, one per constant interval of $F$.
		\item[b)] Let $n$ be a multiple of $m$ plus one, so there exists a $c \in \N$ such that $n = c \cdot m + 1$. It follows that $m$ and $n$ are coprime and $F(D_F) \cap G(D_G) = \emptyset$. Furthermore, for each $k \in \oneto{m-1}$, it holds that $\frac{k}{m} = \frac{c \cdot k}{n-1} \in (\frac{c \cdot k}{n}, \frac{c \cdot k + 1}{n})$. Hence, $\nn{k+1}{F}{G} = \{\tfrac{c \cdot k}{n}, \tfrac{c \cdot k + 1}{n}\}$ for all $k \in \oneto{m-1}$. Since condition (i) in Definition \ref{def:disco}a) is obviously satisfied, Proposition \ref{thm:discoNN}a) states that $F \discoa G$ is equivalent to
		\begin{align*}
			x_{k+1} - x_k &\leq y_{c \cdot k + 1} - y_{c \cdot k} \quad \text{and}\\
			x_{k+1} - x_k &\leq y_{c \cdot k + 2} - y_{c \cdot k + 1}
		\end{align*}
		for all $k \in \oneto{m-1}$. Hence, $2m-2$ comparisons are made overall, two per constant interval of $F$.
		\item[c)] Let the greatest common divisor $d$ of $m$ and $n$ satisfy $1 < d < m$, so there exist $c_F, c_G \in \Nz$ such that $m = c_F \cdot d$ and $n = c_G \cdot d$. Then, $F(D_F) \cap G(D_G) = \{\frac{k}{d}: k \in \oneto{d-1}\}$ since
		\begin{align*}
			F(D_F) \ni \frac{k \cdot c_F}{m} = \frac{k \cdot c_F}{c_F \cdot d} = \frac{k}{d} = \frac{k \cdot c_G}{c_G \cdot d} = \frac{k \cdot c_G}{n} \in G(D_G)
		\end{align*}
		holds for all $k \in \oneto{d-1}$ and because $c_F$ and $c_G$ are coprime by assumption. It follows that $|\nn{\ell+1}{F}{G}| = 1$, if $\ell$ is a multiple of $c_F$, and $|\nn{\ell+1}{F}{G}| = 2$ otherwise. In the former case, the corresponding comparisons (by Proposition \ref{thm:discoNN}a)) have the same structure as in part a), and in the latter case they have the same structure as in part b). Overall, there are $2m-2-(d-1) = (2 c_F - 1) d - 3$ comparisons to be made. Note that the edge cases $d = m$ and $d = 1$ give the situation in part a) and part b), respectively.
	\end{enumerate}
\end{example}

\subsection{Geometric distribution}

Except for the discrete uniform distribution, the geometric distribution is the only popular type of discrete distribution with an explicit representation of the cdf. We use the following version of the geometric distribution: if $X \sim \Geom(\pi)$ with $\pi \in (0, 1)$, then $\Prob(X = k) = \pi \cdot (1-\pi)^{k-1}$ for $k \in \N$. The cdf of $X$ is then given by $F(t) = 1 - (1-\pi)^{\lfloor t \rfloor}$ for $t \geq 0$. Graphically, the dispersion of the distribution seems to decrease as the parameter $\pi$ increases. Furthermore, for $F = \Geom(\pi_F)$ and $G = \Geom(\pi_G)$ with $0 < \pi_G < \pi_F < 1$, $F \sto G$ and even $F <_{st} G$ obviously holds, which already implies $G \not\discoa F$ according to Proposition \ref{thm:discoaInfSto}. The following result gives a sufficient condition for the ordering of two geometric distributions with respect to $\discoa$.

\begin{theorem}
	\label{thm:discoGeom}
	Let $X \sim \Geom(\pi_F)$ and $Y \sim \Geom(\pi_G)$ with $0 < \pi_G < \pi_F < 1$ have cdf's $F$ and $G$. If
	\begin{equation*}
		(\pi_F, \pi_G) \in \left\{ (1 - \lambda^\varrho, 1 - \lambda): \frac{1}{2} < \lambda < 1, \varrho \geq \frac{\log(2\lambda - 1)}{\log(\lambda)} - 1 \right\},
	\end{equation*}
	then $F \discoa G$ holds.
\end{theorem}

\begin{figure}
	\centering{
		\includegraphics[width=\textwidth]{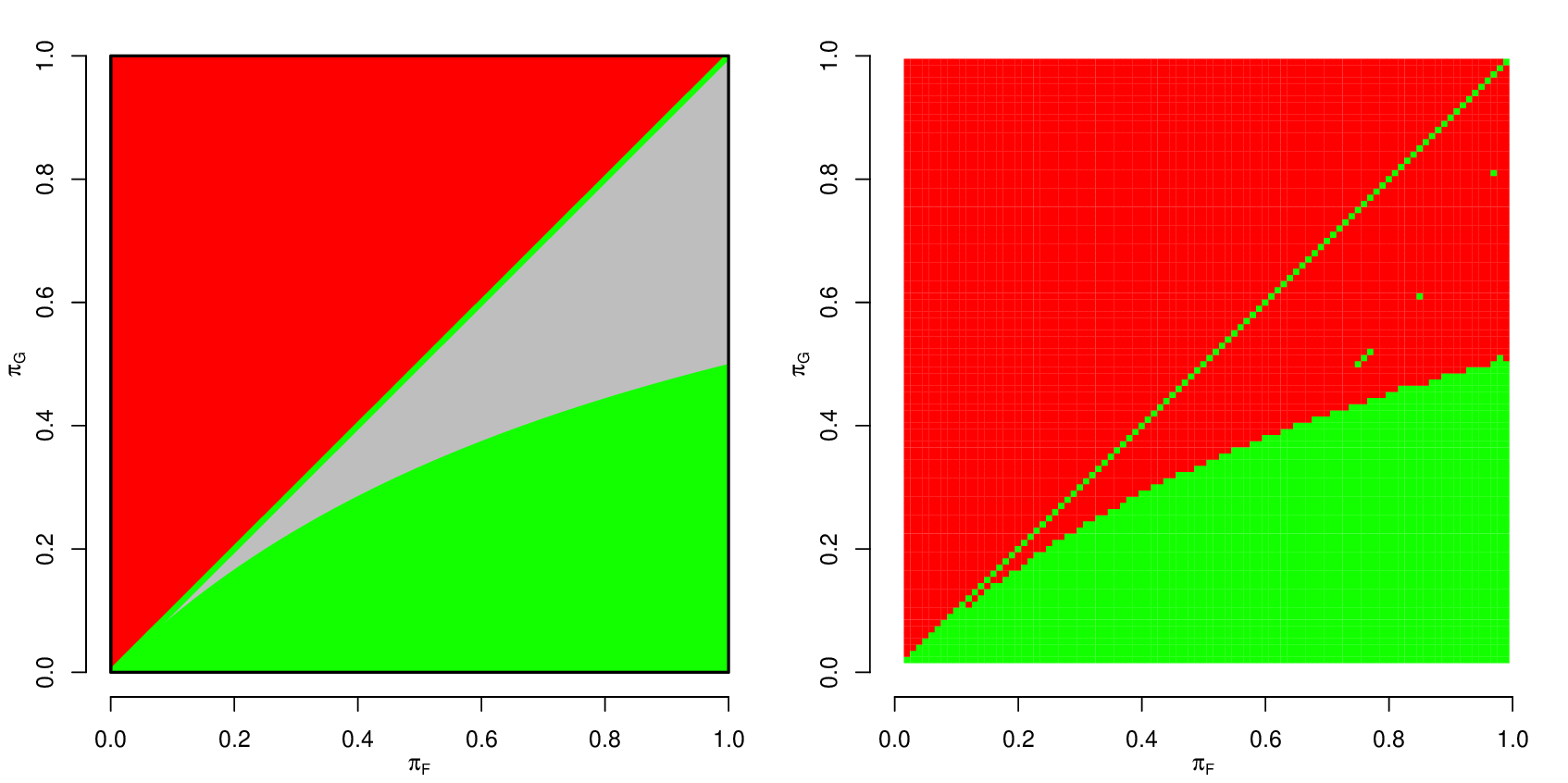}
		\caption{\label{fig:discoGeom}Plot of theoretical (left panel) and numerical (right panel) results concerning $F \discoa G$ for all possible parameter choices $\pi_F, \pi_G \in (0, 1)$ for $F = \Geom(\pi_F)$ and $G = \Geom(\pi_G)$. In green areas, $F \discoa G$ holds; in red areas, $F \not\discoa G$ holds; in grey areas, no result could be obtained.}
	}
\end{figure}

The set of parameter pairs from Theorem \ref{thm:discoGeom} is visualized in the left panel of Figure \ref{fig:discoGeom}, where it is the green area on the lower right. The grey area represents those combinations of parameters, for which no theoretical result could be obtained.
In order to determine the behaviour in these grey areas, a numerical analysis was conducted. Since the support of the geometric distribution is infinite, the cdf's and pdf's were cut off at $10^6$. The results with $0.01$ as increment for the parameters $\pi_F$ and $\pi_G$ are depicted in the right panel of Figure \ref{fig:discoGeom}. The numerical results look almost identical to the theoretical results with the grey area filled in red. It is not clear whether the few sparse green dots in that area actually represent $F \discoa G$ holding or they represent numerical inaccuracies. Either way, the numerical results suggest that the implication in Theorem \ref{thm:discoGeom} is close to being an equivalence as the number of counterexamples for the reverse implication is very small.

\begin{figure}
	\centering{
		\includegraphics[width=\textwidth]{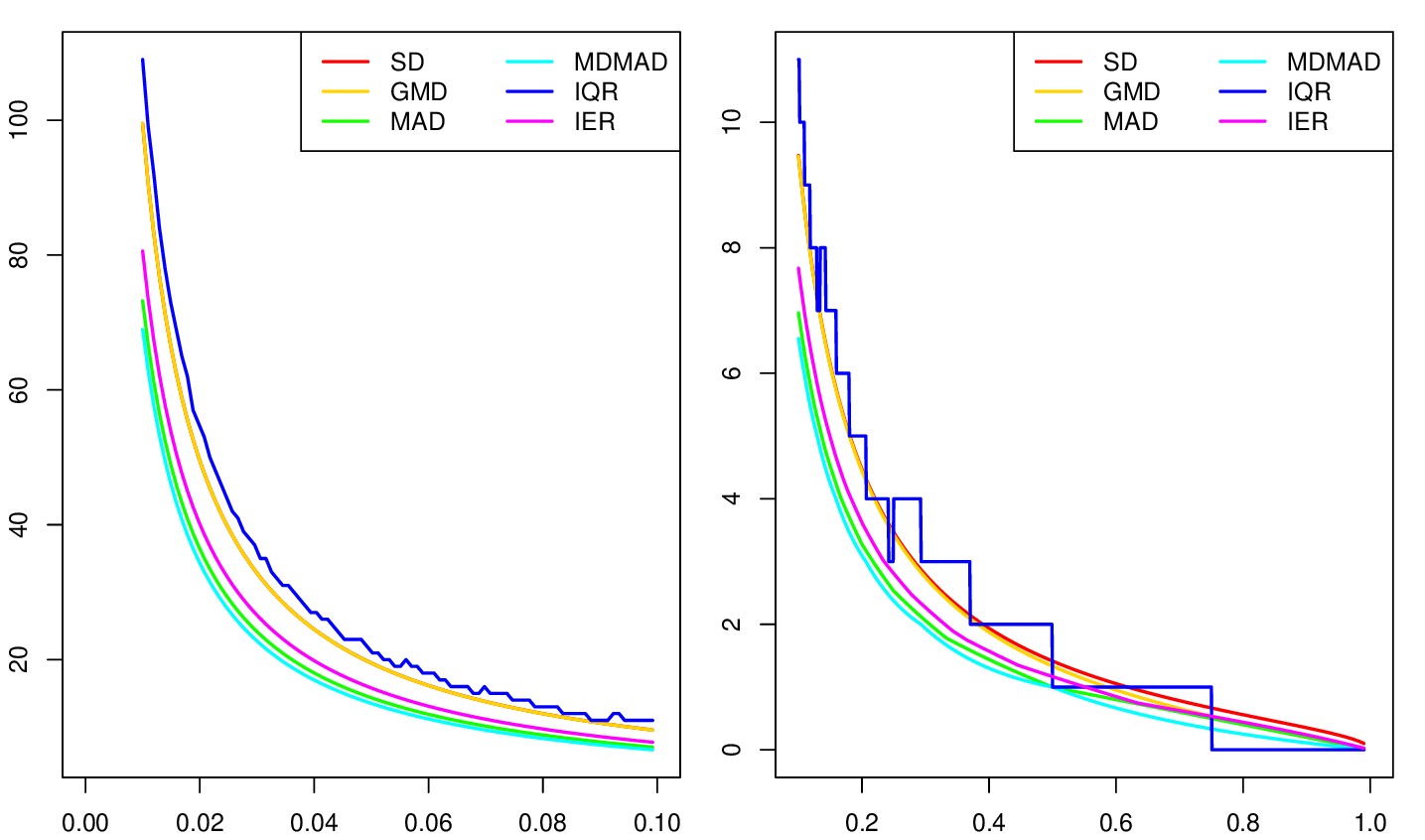}
		\caption{\label{fig:dispMeasGeom}Plot of $\tau(X)$ for six different dispersion measures $\tau$ and $X \sim \Geom(\pi)$, as a function of $\pi \in (0, 1)$. Note the different scales in the two panels.
		}
	}
\end{figure}

The behaviour of the dispersion measures from Section \ref{sec:discDispMeas} applied to geometric distributions is shown in Figure \ref{fig:dispMeasGeom}. First, it is obvious that the graphs all have similar shapes. While that includes the interquartile range $\iqr$ to a certain degree, its graph is the only one that is not decreasing on the entire parameter space. Furthermore, the slopes of all graphs decrease significantly for increasing parameter values. Thus, there is a smaller difference in dispersion between two similarly high values of $\pi$ than there is between two similarly low values of $\pi$. This observation is in agreement with the behaviour of the discrete dispersion order for geometric distributions. Consider the following example: according to Theorem \ref{thm:discoGeom} and Figure \ref{fig:discoGeom}, $F \discoa G$ holds for $\pi_F = 0.15$ and $\pi_G = 0.12$ while $F \not\discoa G$ holds for $\pi_F = 0.9$ and $\pi_G = 0.72$, which differ from each other by the same factor.

\section{Concluding remarks and future research}
\label{sec:conclusion}

As explained in Section \ref{sec:heuristics}, the discrete dispersive orders are carefully constructed based on insights into the original dispersive order, particularly its behavior on discrete distributions. Nevertheless, alternative approaches to defining such an order could also be explored.

One alternative approach could arise out of a small inconsistency of $\discoa$, which can be illustrated using Example \ref{exm:compjao}. Here, part a) is a limit case of part b) if the lengths of some constant intervals converge to zero. However, the requirements for $\discoa$ to hold in the limiting case are quite different to part b). This is due to the fact that, when using $\discoa$, some intervals are not part of any comparisons. An alternative approach that solves this is not restricted to comparing single intervals of $F$ to single intervals of $G$, but the lengths of intervals of the more dispersed $G$ that lie between two intervals of $F$ can be combined. More specifically, the length $x_a - x_{a-1}$ of each constant interval of $F$ is compared with the cumulated length $\sum_{b \in \smi{B}: p_{a-1} \leq q_{b-1} < p_a} (y_b - y_{b-1})$ of all constant intervals of $G$ lower than the next constant interval of $F$. $x_a - x_{a-1}$ is also be compared with the cumulated length $\sum_{b \in \smi{B}: p_{a-2} < q_{b-1} \leq p_{a-1}} (y_b - y_{b-1})$ of constant intervals of $G$ below the interval $[x_{a-1}, x_a)$. This weakens the strong order $\discoa$, but seemingly to a small enough degree that the new order is still meaningful. We conjecture that this new order, just like $\discoo$, is transitive, but not equivalent to $\dispo$ on their joint area of applicability. Thus, it seems like a suitable order to explore in further work. A downside is that it cannot be described with the relation $\stdCompj$ as intuitively as the other proposed discrete orders. Therefore, it might be difficult or even impossible to replicate some of the results in Sections \ref{sec:discDisp} and \ref{sec:discDispMeas}, while the results in Section \ref{sec:specDists} are not going to be improved since they only concern lattice distributions.

Another alternative approach is inspired by the fact that, under suitable regularity conditions, $F \dispo G$ is equivalent to the slopes of their quantile functions being ordered pointwise. This reduces the comparison to a single quantity, eliminating the need to separately consider jump heights and the lengths of constant intervals. The idea would be to interpolate the quantile functions of discrete distributions and then compare their slopes pointwise. However, this approach has several drawbacks.  
First, it fails to compare the first jumps of the distributions in any way. Second, even ignoring this issue, simple examples exhibit counterintuitive behavior -- for instance, consider $X = \frac{1}{2} \tilde{X}$ with $\tilde{X} \sim \Bin(1, \frac{1}{2})$ and $Y = (1 - \varepsilon) \tilde{Y}$ with $\tilde{Y} \sim \Bin(1, 1 - \varepsilon)$ and let $\varepsilon \searrow 0$. Finally, the approach would be highly sensitive to the specific definition of the quantile function used.

\smallskip

The approach presented in this paper has several significant advantages: it establishes a rigorous foundation for discrete dispersion measures that was previously lacking and introduces a new, simple, and effective way to describe and compare discrete distributions. It is informed by the pointwise and local nature of quantile comparisons and aligns these evaluations with the sparse occurrence of probability mass in discrete distributions. Here, the relation $\stdCompj$ plays a pivotal role. Using this relation, the usual stochastic order can also be described in a simple and straightforward manner. Additionally, orders similar to $\discoa$ and $\discoo$ can be derived for higher-order distributional characteristics, such as skewness, which face challenges analogous to those of dispersion in discrete distributions \citep[see][]{ek-zwet}.

Having discussed the merits of this approach based on the relation $\stdCompj$, several open questions remain for future research, particularly regarding the transition from the discrete dispersive orders to the original dispersive order. For instance, consider a continuous cdf $F$ approximated by a sequence of discrete cdf's $F_n$ such that $F_n \stackrel{n \to \infty}{\to} F$ in a given mode of convergence. If the same holds for another continuous cdf $G$, it would be desirable to show that $F_n \discoa G_n$ for all sufficiently large $n$ implies $F \dispo G$. One could also analyze the applicability of the original dispersive order to mixtures of discrete and continuous distributions and explore ways to bridge any gaps in its meaningfulness if necessary.

\section*{Supplementary material}
A supplement to this paper is available, which contains all proofs as well as the application of the proposed discrete dispersion orders to further discrete distributions, adding to Section \ref{sec:specDists}.

\section*{Acknowledgements}
The authors would like to thank two anonymous reviewers for their constructive and helpful comments.

\printbibliography

@article{gerstenberger,
	author = {C. Gerstenberger and D. Vogel},
	title = {On the efficiency of Gini’s mean difference},
	journal = {Statistical Methods \& Applications},
	year = {2015},
	pages = {569--596},
	volume = {24}
}

@article{hurli,
	author = {Werner H{\"u}rlimann},
	title = {On Risk And Price: Stochastic Orderings And Measures},
	journal = {Transactions of the 27th International Congress of Actuaries},
	year = {2002}
}

@article{bellini,
	author = {Fabio Bellini},
	title = {Isotonicity Results for Generalized Quantiles},
	journal = {Statistics \& Probability Letters},
	year = {2012},
	pages = {2017--2024},
	volume = {82}
}

@article{bl3,
	author = {Peter J. Bickel and Erich L. Lehmann},
	title = {Descriptive Statistics for Nonparametric Models. III. Dispersion},
	journal = {The Annals of Statistics},
	volume = {4},
	pages = {1139--1158},
	year = {1976}
}

@article{ek-zwet,
	author = {Andreas Eberl and Bernhard Klar},
	title = {On the Skewness Order of van Zwet and Oja},
	journal = {Mathematical Methods of Statistics},
	volume = {28},
	pages = {262--278},
	year = {2019}
}

@article{ek-patil,
	author = {Andreas Eberl and Bernhard Klar},
	title = {A Note on a Measure of Asymmetry},
	journal = {Statistical Papers},
	volume = {62},
	pages = {1483--1497},
	year = {2021}
}

@article{ek-exlDisp,
	author = {Andreas Eberl and Bernhard Klar},
	title = {Stochastic Orders and Measures of Skewness and Dispersion Based on Expectiles},
	year = {2023},
	journal = {Statistical Papers},
	volume = {64},
	pages = {509--527}
}

@misc{eks-weakDisp,
	author = {Andreas Eberl and Bernhard Klar and Alfonso Su\'arez-Llorens},
	title = {An easily verifiable dispersion order for discrete distributions},
	year = {2025},
	archivePrefix = {"arXiv"}, 
	note = {arXiv:2506.23677},
}

@article{weakDispOrder,
	author = {Alessandra Giovagnoli and H. P. Wynn},
	title = {Multivariate Dispersion Orderings},
	journal = {Statistics \& Probability Letters},
	volume = {22},
	pages = {325--332},
	year = {1995}
}

@article{klar:2010,
	author = {Klar, B. and Petney, T.N. and Taraschewski, H.},
	title = {Quantifying differences in parasite numbers between samples of hosts},
	journal = {Journal of Parasitology},
	volume = {96},
	pages = {856--861},
	year = {2010}
}

@article{landoEmpDispo,
	author = {Tommaso Lando and Idir Arab and Paulo Eduardo Oliveira},
	title = {Transform Orders and Stochastic Monotonicity
of Statistical Functionals},
	journal = {Scandinavian Journal of Statistics},
	doi = {10.1111/sjos.12629},
	    note = {Early view},
	year = {2022}
}

@book{mueller-stoyan,
	author = {Alfred M{\"u}ller and Dietrich Stoyan},
	title = {Comparison Methods for Stochastic Models and Risks},
	publisher = {Wiley},
	series = {Wiley Series in Probability and Statistics},
	location = {Chichester},
	year = {2002}
}

@phdthesis{Munderle:2005,
	author = {Münderle, M.},
	title = {Ökologische, morphometrische und genetische Untersuchungen an Populationen des invasiven Schwimmblasennematoden Anguillicola crassus aus Europa und Taiwan},
    school = {University of Karlsruhe},
	year = {2005}
}

@article{Munderle:2006,
	author = {Münderle, M. and Taraschewski, H. and Klar, B. and Chang, C.W. and Shiao, J.C. and Shen, K.N. and He, J.T. and Lin, S.H. and Tzeng, W.N.},
	title = {Occurrence of Anguillicola crassus (Nematoda: Dracunculoidea) in Japanese eels Anguilla japonica from a river and an aquaculture unit in southwest Taiwan},
	journal = {Diseases of Aquatic Organisms},
	volume = {71},
	pages = {101--108},
	year = {2006}
}

@article{oja,
	author = {Hannu Oja},
	title = {On Location, Scale, Skewness and Kurtosis of Univariate Distributions},
	journal = {Scandinavian Journal of Statistics},
	volume = {8},
	pages = {154--168},
	year = {1981}
}

@article{gmdDiloOld,
	author = {H{\'e}ctor M. Ramos and Miguel A. Sordo},
	title = {Dispersion Measures and Dispersive Orderings},
	journal = {Statistics \& Probability Letters},
	volume = {861},
	pages = {123--131},
	year = {2003}
}

@book{shaked-sh,
	author = {Moshe Shaked and J. George Shanthikumar},
	title = {Stochastic Orders},
	series = {Springer Series in Statistics},
	publisher = {Springer},
	year = {2006}
}

@article{gmdDiloNew,
	author = {Miguel A. Sordo and Marilia C. de Souza and Alfonso Su{\'a}rez-Llorens},
	title = {Testing Variability Orderings by Using Gini's Mean Differences},
	journal = {Statistical Methodology},
	volume = {32},
	pages = {63--76},
	year = {2016}
}

@booklet{zwet,
	author = { {v}an Zwet, W. R.},
	shortauthor = { van Zwet, W. R.},
	sortname = {Zwet, W. R. van},
	title = {Convex Transformations of Random Variables},
	location = {Mathematical Centre Tracts, Amsterdam},
	year = {1964}
}

@article{IER,
	author = {Fabio Bellini and Lorenzo Mercuri and Edit Rroji},
	title = {Implicit Expectiles and Measures of Implied Volatility},
	journal = {Quantitative Finance},
	year = {2018},
	pages = {1851--1864},
	volume = {18}
}

@article{bl2,
	author = {Peter J. Bickel and Erich L. Lehmann},
	title = {Descriptive Statistics for Nonparametric Models. II. Location},
	journal = {The Annals of Statistics},
	year = {1975},
	pages = {1045--1069},
	volume = {3}
}

@inbook{bl4,
	author = {Peter J. Bickel and Erich L. Lehmann},
	title = {Descriptive Statistics for Nonparametric Models. IV. Spread},
	booktitle = {Contributions to Statistics. Jaroslav Hajek Memorial Volume},
	year = {1979},
	pages = {33--40},
	publisher = {D. Reidel Pub. Co.}
}

@article{variance_hist,
	author = {Michael Kourkoulos and Constantinos Tzanakis},
	title = {History, and Students'  Understanding of Variance In Statistics},
	journal = {BSHM Bulletin},
	year = {2010},
	pages = {168--178},
	volume = {25}
}

\includepdf[pages=-]{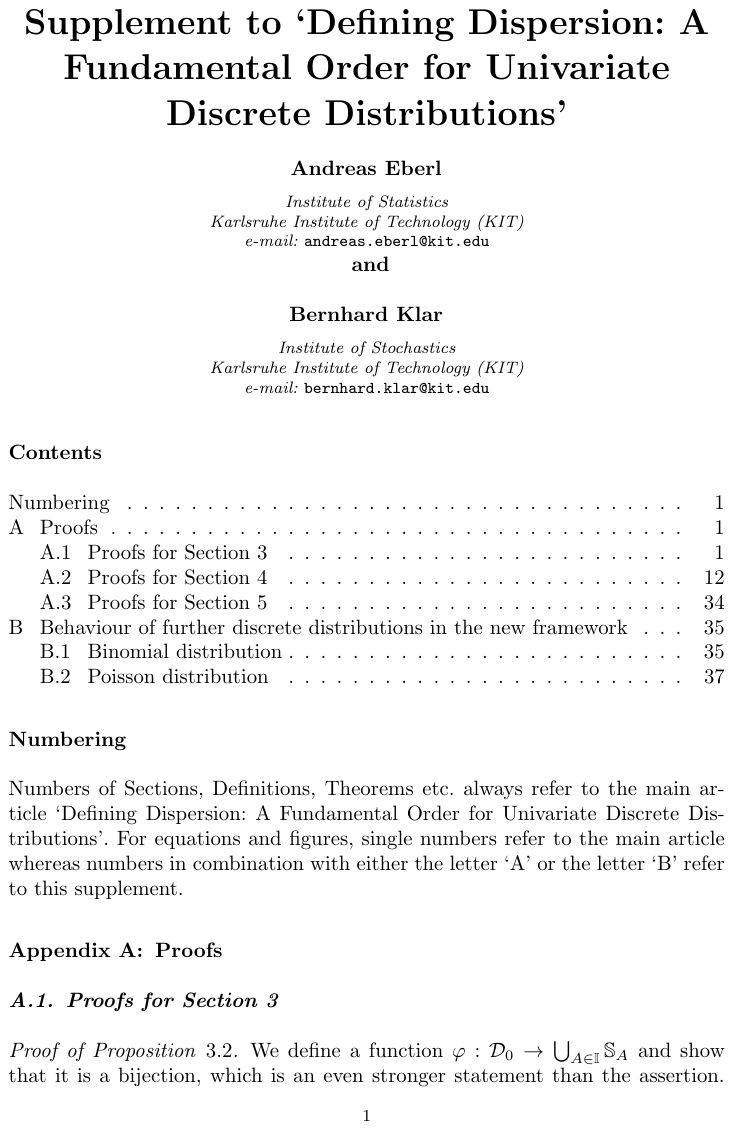}

\end{document}